\documentclass[a4paper,twocolumn,11pt,accepted=2025-07-01]{quantumarticle}
\pdfoutput=1

\usepackage[utf8]{inputenc}
\usepackage[english]{babel}
\usepackage[T1]{fontenc}
\usepackage{amsmath}

\usepackage{hyperref}
\usepackage{lipsum}

\usepackage{cite}

\usepackage{amssymb,amsmath, amsthm}
\usepackage{graphicx}
\usepackage{subfigure}

\usepackage{braket}
\usepackage{xcolor}
\usepackage[title]{appendix}

\def\comment#1{}
\def\bra#1{\mathinner{\langle{#1}|}}
\def\ket#1{\mathinner{|{#1}\rangle}}

\def\l{\left}
\def\r{\right}

\def\comment#1{}

\begin{document}
	
	\title{Quantum-Enhanced Sensitivity through Many-Body Bloch Oscillations}
	
	\author{Hassan Manshouri}
	\email[]{h.manshouri@ph.iut.ac.ir}
	\affiliation{Department of Physics, Isfahan University of
		Technology, Isfahan 84156-83111, Iran}
	
	\author{Moslem Zarei}
	\email[]{m.zarei@iut.ac.ir}
	\affiliation{Department of Physics, Isfahan University of Technology, Isfahan 84156-83111, Iran}
	
	\author{Mehdi Abdi}
	\email[]{mehabdi@gmail.com}
	\affiliation{Wilczek Quantum Center, School of Physics and Astronomy, Shanghai Jiao Tong University, Shanghai 200240, China}
	
	\author{Sougato Bose}
	\email[]{s.bose@ucl.ac.uk}
	\affiliation{Department of Physics and Astronomy, University College London, Gower Street, WC1E6BT, London, United Kingdom}
	
	\author{Abolfazl Bayat}
	\email[]{abolfazl.bayat@uestc.edu.cn}
	\affiliation{Institute of Fundamental and Frontier Sciences, University of Electronic Science and Technology of China, Chengdu 611731, China}
	\affiliation{Key Laboratory of Quantum Physics and Photonic Quantum Information, Ministry of Education,
		University of Electronic Science and Technology of China, Chengdu 611731, China}
		
	\maketitle
	
\begin{abstract}
We investigate the sensing capacity of non-equilibrium dynamics in quantum systems exhibiting Bloch oscillations. By focusing on the resource efficiency of the probe, quantified by quantum Fisher information, we find different scaling behaviors in two different phases, namely localized and extended. Our results provide a quantitative ansatz for quantum Fisher information in terms of time, probe size, and the number of excitations. In the long-time regime, the quantum Fisher information is a quadratic function of time, touching the Heisenberg limit. The system size scaling drastically depends on the phase changing from quantum-enhanced scaling in the extended phase to size-independent behavior in the localized phase. Furthermore, increasing the number of excitations always enhances the precision of the probe, although, in the interacting systems the enhancement becomes less eminent than the non-interacting probes. This is due to the induced localization by increasing the interaction between the excitations. We show that a simple particle configuration measurement together with  a maximum likelihood estimation can closely reach the ultimate precision limit in both single- and multi-particle probes.
\end{abstract}
\linebreak[4]
	
\section{Introduction}%
The delicate nature of quantum systems makes them naturally suitable for sensing gravitational, magnetic, and electric fields with unprecedented precision well beyond the capacity of classical probes~\cite{degen2017quantum}. 
In order to estimate an unknown parameter $h$, encoded in the quantum state of a probe $\rho(h)$, one has to perform measurement. For a given measurement, described by Positive Operator-Valued Measure operators $ \{ \pi_x\}$ (POVM), the results follow a classical probability distribution in which any outcome appears with the probability $p_x(h){=}\mathrm{Tr}[\pi_x \rho(h)]$. By post-processing the measurement results, one can construct an estimator $\hat{h}$. The estimator has to be asymptotically unbiased in the sense that its expectation value gives the real value of the unknown parameter, namely $\langle \hat{h}\rangle{=}h$, in the limit of large measurement samples.
In this situation, the precision of the probe for sensing parameter $h$ is quantified by the standard deviation of the estimator, which we call it  $\delta h$. For a given measurement $\{ \pi_x \}$ and an unbiased estimators, the precision $\delta h$ is asymptotically  bounded by  Cram\'{e}r-Rao inequality 
$\delta h {\ge} 1/\sqrt{\mathcal{M}\mathcal{F}_C}$, where $\mathcal{M}$ is the number of samples and $ \mathcal{F}_{C} $  is called Classical Fisher information (CFI) 
which is defined as~\cite{fisher1922mathematical,paris2009quantum,meyer2021fisher,Liu_2019}
\begin{equation}\label{CFI}
     \mathcal{F}_{C}(h)= \sum_{x} \frac{1}{p_x(h) }\left(\frac{d p_x}{dh}\right)^2.
\end{equation}
The estimator that  can saturate the above inequity in the asymptotic limit is called ``efficient'' estimator. Therefore, in order to saturate the Cram\'er-Rao inequality of Eq.~(\ref{CFI}) one has to rely on an optimal estimator which is both unbiased and efficient. 
Bayesian and maximum likelihood are known for being asymptotically optimal estimators. 
According to Laplace-Bernstein–von Mises theorem~\cite{berger2013statistical,lehmann2006theory}, in the limit of large repeating measurement samples, the Bayesian and maximum likelihood converge to the same value of estimation. In particular, estimation by a Bayesian estimator with uniform prior in large repetitions is literally identical to maximum likelihood.

Furthermore, one can go one step further and try to find a precision bound which is independent of measurement basis. To accomplish this, one can maximize the CFI with respect to all possible 
POVMs
to obtain Quantum Fisher Information (QFI), namely $\mathcal{F}_Q{=}\underset{\{ \pi_x\}}{\mathrm{max}}\mathcal{F}_C$. The QFI indeed describes the ultimate precision limit for which the Cram\'{e}r-Rao inequality can be written as 
\begin{equation}
    \delta h \ge \frac{1}{\sqrt{\mathcal{M}\mathcal{F}_C} } \ge \frac{1}{\sqrt{\mathcal{M}\mathcal{F}_Q}}.
\end{equation}
While the saturation of the first inequality requires an optimal estimator, the saturation of the second inequality demands both optimized measurement and estimator. Interestingly, the maximization over all possible measurements in the definition of the QFI can be reduced to a closed formula~\cite{paris2009quantum}, see Refs.~\cite{Liu_2019,meyer2021fisher} for detailed discussions on QFI. For instance, in the case of pure states $\rho(h){=}\ket{\Psi(h)}\bra{\Psi(h)}$ the QFI takes the following form~\cite{paris2009quantum}
\begin{equation}
    \mathcal{F}_Q(h)= 4 [\braket{\partial_h \Psi|\partial_h \Psi}-|\braket{\partial_h \Psi|\Psi}|^2].
\end{equation}
The performance of a sensor is determined by the scaling of its QFI with respect to resources (probe size and time), namely $\mathcal{F}_Q {\sim} L^\beta$, where $L$ is the resource and $\beta$ is an exponent. In the absence of quantum features, 
one can at best achieve $\beta{=}1$ (i.e. standard limit). On the other hand, by properly harnessing quantum features, such as entanglement,  the precision might be enhanced to $\beta >1$, known as quantum-enhanced sensitivity, in which a special case is $\beta{=}2$ (i.e. Heisenberg limit)~\cite{giovannetti2004quantum}.

Quantum criticality is known to be instrumental for achieving quantum-enhanced sensitivity~\cite{Montenegro_2025}. In fact, various forms of criticality have been exploited for sensing purposes, including first-order~\cite{Sarkar_2025,raghunandan2018high,heugel2019quantum,yang2019engineering}, second-order~\cite{zanardi2006ground,zanardi2007mixed,gu2008fidelity,zanardi2008quantum,invernizzi2008optimal,gu2010fidelity,gammelmark2011phase,skotiniotis2015quantum,rams2018limits,wei2019fidelity,chu2021dynamic,liu2021experimental,montenegro2021global,mirkhalaf2021criticality,di2021critical,Salvia2023Critical,garbe2022exponential}, dissipative~\cite{fernandez2017quantum,baumann2010dicke,baden2014realization,klinder2015dynamical,rodriguez2017probing,fitzpatrick2017observation,fink2017observation,ilias2022criticality,Ilias2023Criticality,Alipor2014Quantum}, topological~\cite{budich2020non,sarkar2022free,koch2022quantum,yu2022experimental}, Floquet~\cite{mishra2021driving,mishra2022integrable}, time crystals~\cite{montenegro2023quantum,yousefjani2024discrete,iemini2023floquet}, non-Hermitian systems~\cite{Wiersig2014Enhancing, Liu2016Metrology, Langbein2018No, Lau2018Fundamental, Zhang2019Quantum, Chen2019Sensitivity}, and Stark~\cite{he2023stark,yousefjani2023Long,yousefjani2024nonlinearityenhanced} phase transitions.  However, despite a few attempts for realizing criticality-based sensors~\cite{ding2022enhanced, liu2021experimental}, their experimental implementation faces real challenges. 
Normally, in most of such sensors, the probe should be initialized in one of its eigenstates, e.g. the ground state, which is very challenging in practice and may require extreme cooling or adiabatic state preparation whose time resource scales undesirably with the system size~\cite{chu2021dynamic}. In addition, the parameter interval over which such criticality-based probe achieves quantum-enhanced sensitivity is often very narrow covering only around the phase transition point. A possible solution for addressing these challenges is to go beyond equilibrium physics and exploit non-equilibrium dynamics of many-body systems~\cite{balatsky2024quantum,bhattacharyya2024even, bhattacharyya2024restoring, montenegro2022sequential,yang2023extractable}, which  are easy to implement in various physical platforms~\cite{rispoli2019quantum,franke2023quantum, mi2022noise, gong2021quantum, kohlert2023exploring}. 
Such systems do not require complex initialization and may allow quantum-enhanced sensitivity over a wider range. Several open questions exist, including: (i) what types of non-equilibrium systems may lead to quantum-enhanced sensitivity? and (ii) can such probes operate optimally across an entire phase of matter rather than a narrow region around the phase transition point?

Bloch oscillation is a fundamental phenomenon in condensed matter physics in which a particle under the impact of a gradient potential (i.e. a constant force) oscillates in a regular lattice~\cite{bloch1929quantenmechanik}. The gradient potential naturally makes the neighboring lattice sites off-resonant and hence suppresses the tunneling of particles. Unlike classical systems in which particles simply move in the direction of the external force, this suppression of tunneling in quantum systems induces Wannier-Stark localization in the system~\cite{wannier1960wave}. The localization properties of Wannier-Stark systems have been extensively studied in both theory~\cite{fukuyama1973tightly,holthaus1995random,kolovsky2003bloch,kolovsky2008interplay,kolovsky2013wannier,van2019bloch,schulz2019stark,wu2019bath,bhakuni2020drive,bhakuni2020stability,yao2020many,chanda2020coexistence,taylor2020experimental,wang2021stark,zhang2021mobility,yao2021many,doggen2022many,zisling2022transport,burin2022exact,bertoni2024local,lukin2022many,vernek2022robustness,Doggen2021Stark,Sahoo2024Localization} and experiments~\cite{morong2021observation,preiss2015strongly,kohlert2021experimental,karamlou2022quantum}. Bloch oscillations have also been observed in semiconductor devices~\cite{leo1992observation},  optical waveguides~\cite{peschel1998optical,jiang2023waveguide,tang2018waveguide}, cold atoms in optical lattices~\cite{dahan1996bloch} and superconducting simulators~\cite{guo2021observation,guo2021stark}. Unlike the conventional second-order quantum phase transitions which only affect the ground state, 
	the  Wannier-Stark localization takes place across the entire spectrum. This implies that the impact of such transitions should be observable in non-equilibrium dynamics, where many eigenstates are involved. Recently, it has been shown that individual eigenstates of the Wannier-Stark systems can be used for sensing the gradient field with quantum-enhanced precision~\cite{he2023stark}. A natural question is whether one can exploit experimental-friendly non-equilibrium dynamics of such systems for sensing purposes. 
	
In this paper, we explore the sensing capacity of the Bloch oscillations in single- and many-body Stark systems. We provide a comprehensive analysis for the scaling of the QFI in terms of time, probe size, and the number of excitations. We show that indeed many-body systems with Bloch oscillations may allow for sensing precision with quantum-enhanced scaling. Unlike critical probes at equilibrium we do not demand complex initialization and the probe operates optimally over the entire extended phase. We also demonstrate that in a practical setup a simple particle configuration measurement together with a maximum likelihood estimation can closely saturate the bound given by the Cram\'{e}r-Rao inequality.

	\section{Bloch oscillation in Stark systems}%
\begin{figure} \label{F}
		\includegraphics[width=.49\linewidth]{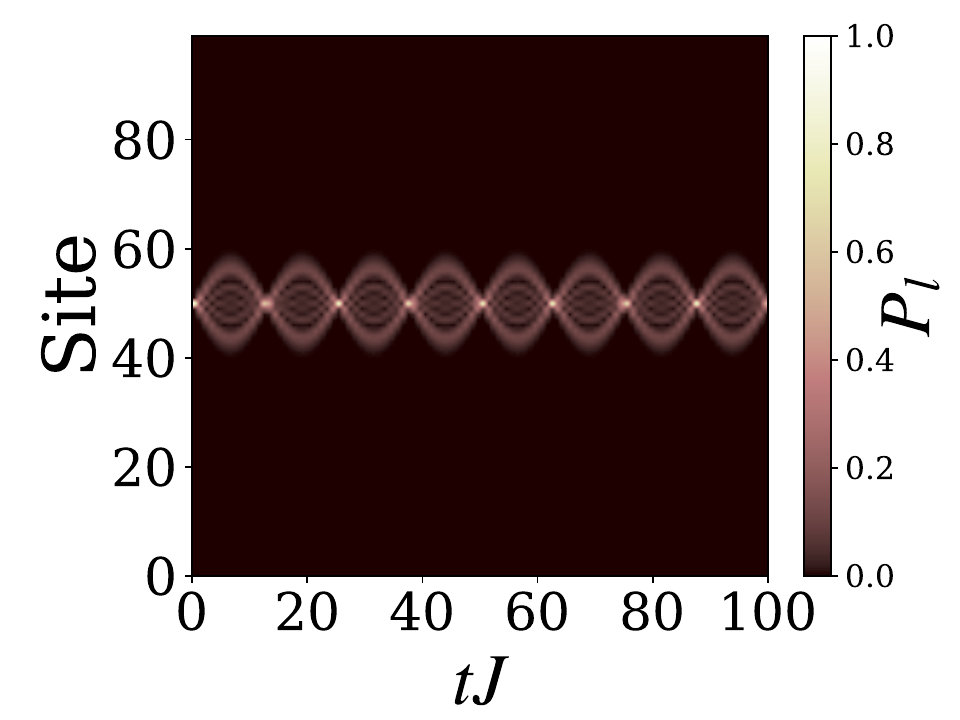}	
		\put(-90,70){\color{white}(a)}
		\includegraphics[width=.49\linewidth]{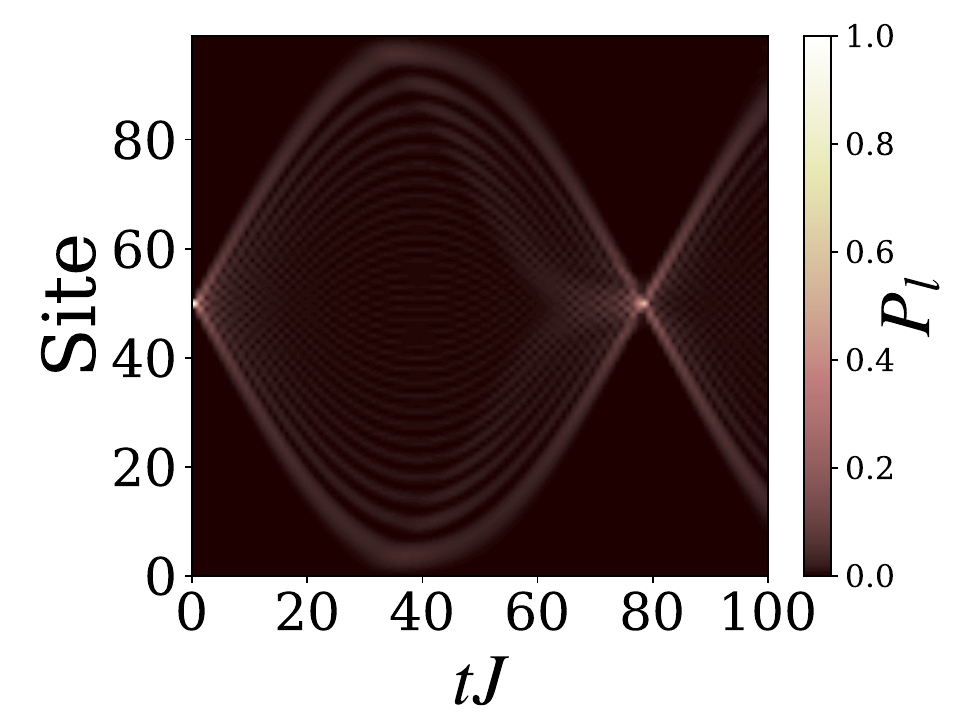}
		\put(-90,70){\color{white}(b)}\\
		\includegraphics[width=.49\linewidth]{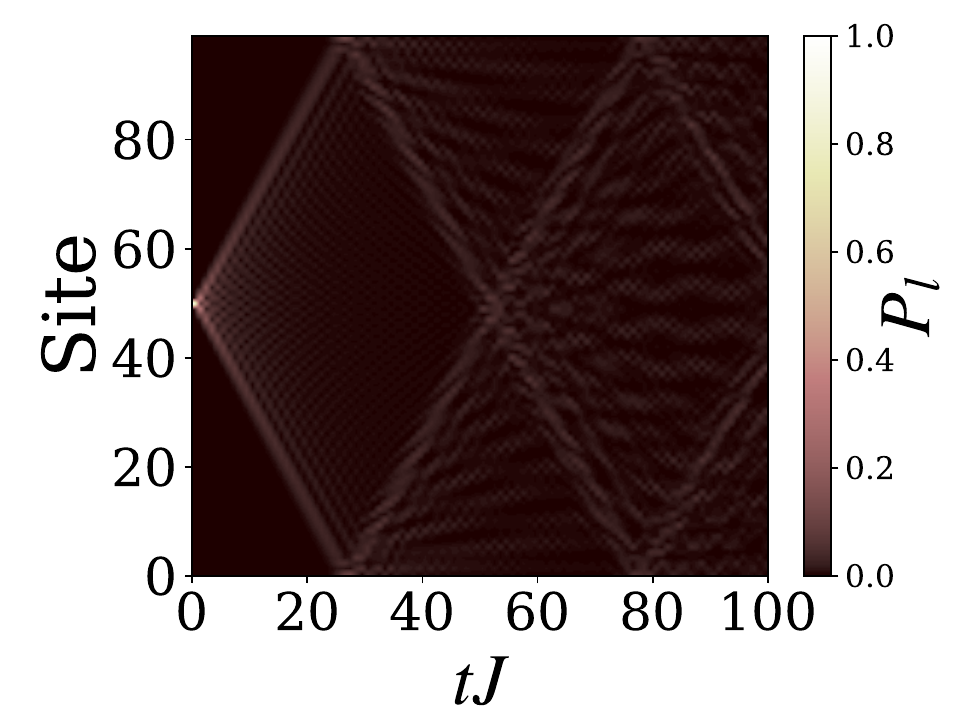}
		\put(-90,70){\color{white}(c)}
		\includegraphics[width=.49\linewidth]{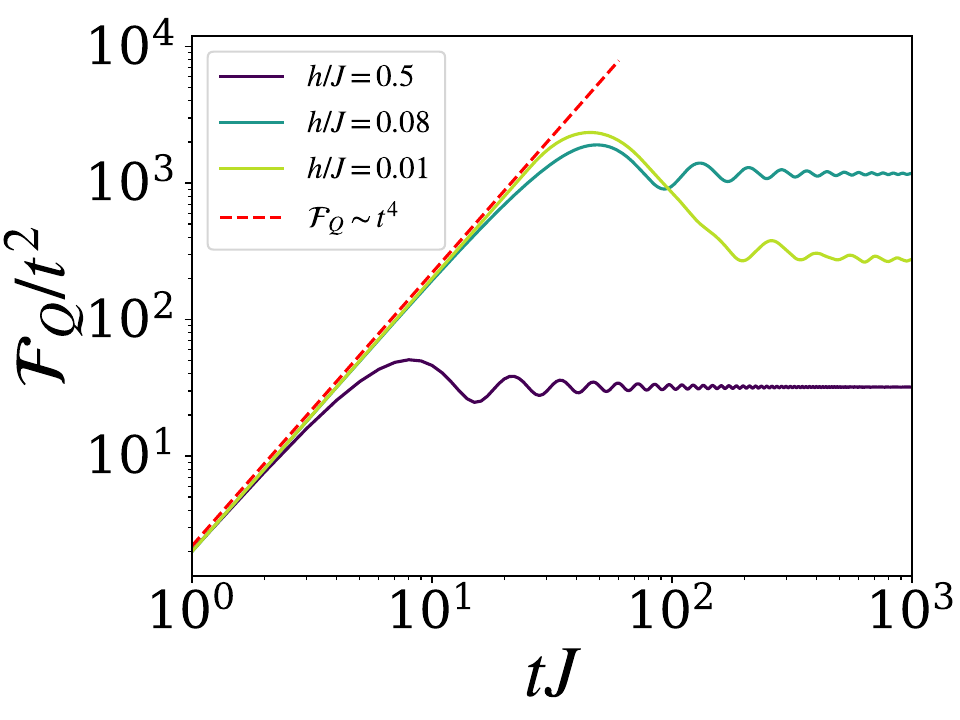}
		\put(-20,23){\small(d)}
		\caption{The time evolution of the population $P_l{=}|\braket{l|\Psi(t)}|^2$ for a system with size $L{=}100$, initially localized at $l{=}50$, for different phases: (a) $h/J=0.5$ in the localized phase, (b) $h/J{=}0.08$ at the transition point, (c) $h/J{=}0.01$ in extended regime. (d) the time evolution of the normalized QFI for different values of $h$ which shows that in the long-time limit, the QFI scales quadratically with time, while at the short-time limit, it scales with time as $\mathcal{F}_Q \sim t^4$.}
		\label{fig:3dp0}
	\end{figure}

	We establish our theory with the simple case of a single excitation in a one-dimensional lattice of size $L$  described by the Hamiltonian ($\hbar{=}1$)
	\begin{equation}\label{Hstark}
		H = -J \sum_{l=1}^{L-1} \ket{l}\!\bra{l+1}+\ket{l+1}\!\bra{l} + h \sum_{l=1}^{L} l \ket{l}\!\bra{l}~,
	\end{equation} 
	where $J$ is the exchange coupling, $h$ is the gradient field and 
	$\ket{l}$ represents the excitation at site $l$. 
	The gradient field induces a position-dependent energy shift on each site which suppresses the particle tunneling. This leads to the localization of the wave function into a limited number of sites, known as Wannier-Stark localization~\cite{wannier1960wave}. 
	In the limit of $L {\rightarrow} \infty$, the eigenstates of Hamiltonian~\eqref{Hstark}, the Wannier-Stark states, are found as
        $\ket{E_m}{=}\sum_{l=1}^{L} \mathcal{J}_{l-m}\big(\frac{2J}{h}\big) \ket{l}$
		whose corresponding eigenvalues are $E_m{=}m h$~\cite{fukuyama1973tightly}.
	Here, $\mathcal{J}_l(\cdot)$ are the Bessel function of the first kind.
	The state $\ket{E_m}$ is a superposition of the states centered at the $m$th site with the spreading width
	about $4J{/}h$ lattice periods~\cite{Holthaus1996}. 
	Any initial state $\ket{\Psi(0)}{=}\sum_l f_l\ket{l}$, evolves under the action of the Hamiltonian $\ket{\Psi(t)}{=}e^{-iHt}\ket{\Psi(0)}$ which can be cast into
		$\ket{\Psi(t)}{=}\sum_{l,l'=1}^{L}\mathcal{K}_{l,l'}(t)f_{l'} \ket{l}$
	with the evolution  propagator $\mathcal{K}_{l,l'}(t){=}\sum_{m=1}^{L} \braket{l|E_m}\!\braket{E_m|l'}e^{-iE_m t}$~\cite{fukuyama1973tightly}. Note that in the large $L$ limit,  
	the energies are equidistant, i.e. $\Delta E {\equiv} E_{m+1}{-}E_m {=} h$. Hence, the evolution propagator exhibits a time-periodic behavior with the characteristic time $T_{\text{Bloch}}{=}2 \pi /h$, which leads to the Bloch oscillations. In order to see the dynamical behavior of the system, we depict $P_l(t){=}|\langle{l}|\Psi(t)\rangle|^2$ as a function of time in Figs.~\ref{fig:3dp0}(a)-(c) for three different values of $h$ in a system of size $L{=}100$,  initialized at the central site. For larger values of $h$, shown in Fig.~\ref{fig:3dp0}(a),
	one clearly observes the periodic spreading of the excitation over a finite distance and returning to its original site. Here, the Bloch oscillations indicate a localized phase.
	By decreasing $h$ the extent of periodic wave packet spreading increases until $h{=}h_{c} {\simeq} 8J{/}L$, at which the wave packet delocalizes across the entire system and then localizes back, see Fig.~\ref{fig:3dp0}(b). This is the point of transition in which the localized phase transforms into an extended phase. Indeed, by further decreasing of $h$, a new behavior emerges as the excitation spreads across the entire system but does not fully return to its initial localized site, signaling an ergodic behavior~\cite{Holthaus1996, he2023stark}.
	
\section{Model 1: Single excitation probe}%

A direct consequence of the above evolution is to imprint the information of $h$ in the quantum state $\ket{\Psi(t)}$, which is clearly evidenced in Figs.~\ref{fig:3dp0}(a)-(c). 
In Fig.~\ref{fig:3dp0}(d), we plot the normalized QFI, namely $\mathcal{F}_Q/t^2$, as a function of time for three values of $h$. One notices a rapid growth in time and eventually saturation indicating that after a transient time, the QFI eventually scales quadratically, i.e. $\mathcal{F}_Q {\sim} t^2$. 
Crucially, for a given length we observe that the highest long-time value of $\mathcal{F}_Q/t^2$  is attained when the system is tuned close to its transition point $h{=}h_c$.
	Indeed, this is the regime that exhibits the widest sustainable superposition of the wave packet $\ket{\Psi(t)}$. 
	Also, Fig~\ref{fig:3dp0}(d) shows the size-independent scaling in early times of the QFI as $\mathcal{F}_Q {\sim} t^4$.
	
	\begin{figure}
		\includegraphics[width=1\linewidth]{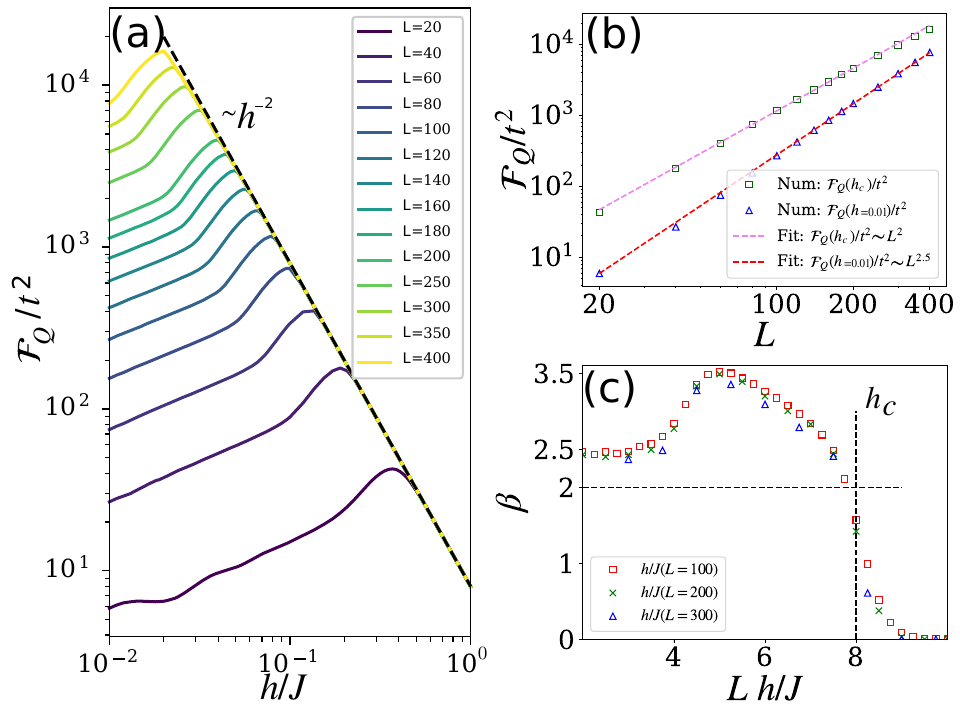}
		\caption{(a) Long-time behavior of $\mathcal{F}_{Q}/t^2$ as a function of $h/J$  for different system sizes.  (b) The scaling of the long-time limit of $\mathcal{F}_{Q}/t^2$ with respect to the system size $L$ at the transition point $h{=}h_c$ (green squares) and deep into the extended phase  $h/J{=}0.01$ (blue triangles). The dashed lines show the best fitting functions. 
			(c) Exponent $\beta$  as a function of $Lh/J$ for various system sizes $L$. The transition point is specified at $h_c{=}8J/L$ whose corresponding exponent is $\beta{=}2$.
		}
		\label{fig:2}		  
	\end{figure}
	
After investigating the time dependence of the QFI, we now focus on scaling with respect to the probe size $L$. In Fig.~\ref{fig:2}(a) we plot the long-time values of $\mathcal{F}_Q/t^2$ as a function of $h$ for different system sizes.
The localization transition points for different system sizes $L$ are clearly identified, which happens around $h_{c} {\simeq} 8J/L$. Note that this is the point at which the Bloch oscillations cover the entire system.
	In the localization phase, $\mathcal{F}_{Q}$ becomes size-independent and scales inversely with square of the gradient field.
	On the other hand, the QFI becomes size-dependent for $h \le h_{c}$.
	These suggest that
	\begin{align}
		\mathcal{F}_{Q} &\sim h^{-2} t^2 L^0,  \;\; \text{   for } h>h_c~, \nonumber\\
		\mathcal{F}_{Q} &\sim  t^2 L^\beta, \hspace{7.5mm}   \text{   for } h \le h_{c}~,
		\label{eq:QFI_Loc}
	\end{align}
where the scaling exponent $\beta$ may vary with $h$.
Indeed, at the transition point $h_{c} {\simeq} 8J/L$, the QFI from both sides of the transition point, given in Eqs.~\eqref{eq:QFI_Loc}, match with each other which results in $\beta{=}2$. 
This is confirmed numerically as for different system sizes we find maximized $\mathcal{F}_Q/t^2$ with respect to $h$ at large time $t$ where $\mathcal{F}_Q/t^2$ is saturated, see the green squares in Fig.~\ref{fig:2}(b). 
In the extended phase, the scaling becomes even better as for $h/J{=}0.01$ the exponent $\beta$ increases to $\beta \simeq 2.5$, depicted by blue triangles in Fig.~\ref{fig:2}(b). Finally, to better understand the metrological aspects of our model we extract the scaling exponent $\beta$ for arbitrary values of $h$, see Appendix \ref{A} for more details. The results are shown in Fig.~\ref{fig:2}(c), where one can see a clear quantum-enhanced scaling with $\beta>2$ in the entire extended phase. The maximum scaling is obtained around $hL/J {\simeq} 5$ whose exponent is $\beta {\simeq} 3.5$. As mentioned before, at the transition point $h{=}h_c$ the QFI scales with Heisenberg precision, i.e. $\beta{=}2$, which then drops to $\beta{=}0$ in the localized phase. It is worth emphasizing that the region below the critical point, where $\beta {>}2$ is a region which only exit is finite system sizes and its width shrinks by increasing the length of the chain. Therefore, at the large $L$ limit, where $h_c{\rightarrow} 0$, the obtainable exponent would be $\beta{\le} 2$.

The above discussion and the studies in the following sections rely on the scaling of $\mathcal{F}_Q/t^2$ at long times. However, we have to specify what long time means. As mentioned above, in the localized phase the Bloch oscillations occur with the period of $T_{\mathrm{Bloch}}=2\pi/h$. By careful checking of the evolutions in Fig.~\ref{fig:3dp0}(d), one can see that $\mathcal{F}_Q/t^2$ saturates only after a few ($\sim 5$) oscillations. By decreasing $h$ the system becomes less localized and more extended which makes the saturation time longer. Nonetheless, the saturation time in the units of $T_{\mathrm{Bloch}}$ decreases. The behavior remains the same in the extended phase such that the saturation takes place after $t{=}2\pi/h$.  Therefore, generally speaking, the long-time behavior can be obtained after a few ($\sim 1-5$) multiples of $t{=}2\pi/h$ for any values of the gradient field $h$.

	\section{Model 2: many-body probe (half filling)}\label{MBp}%
\begin{figure}
		\includegraphics[width=.49\linewidth]{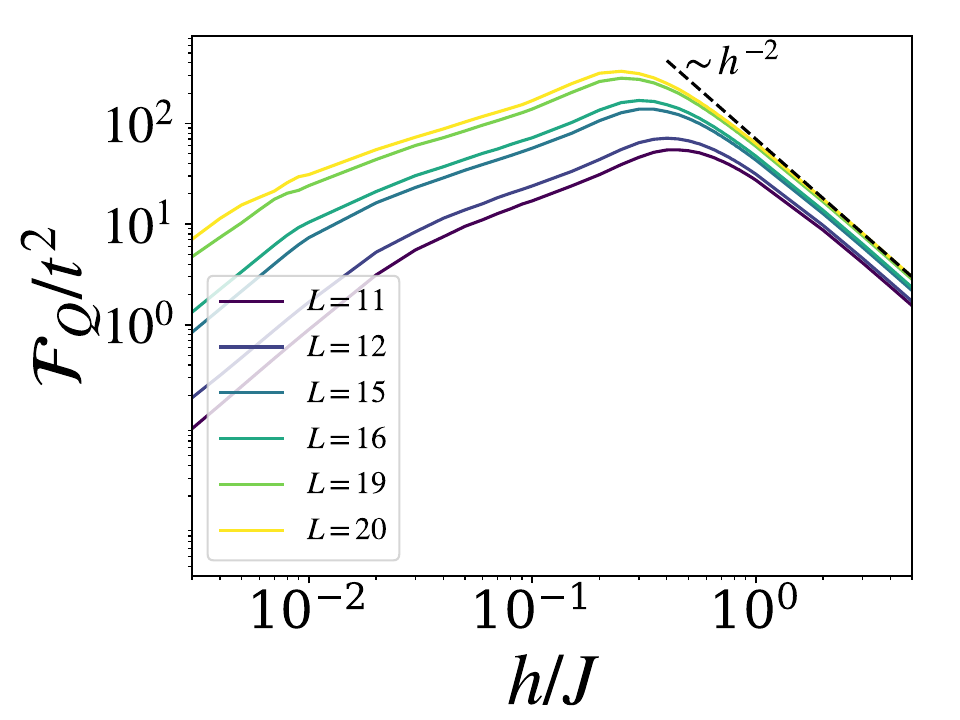}	
		\put(-90,70){(a)}	
		\includegraphics[width=.49\linewidth]{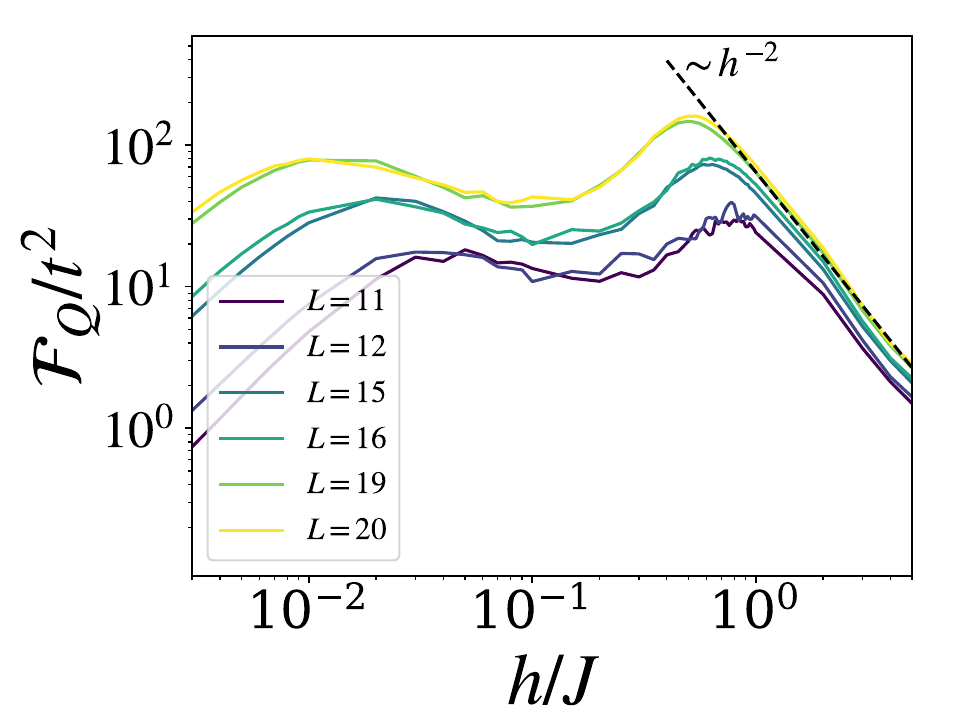}
		\put(-90,70){(b)} \\
		\includegraphics[width=.49\linewidth]{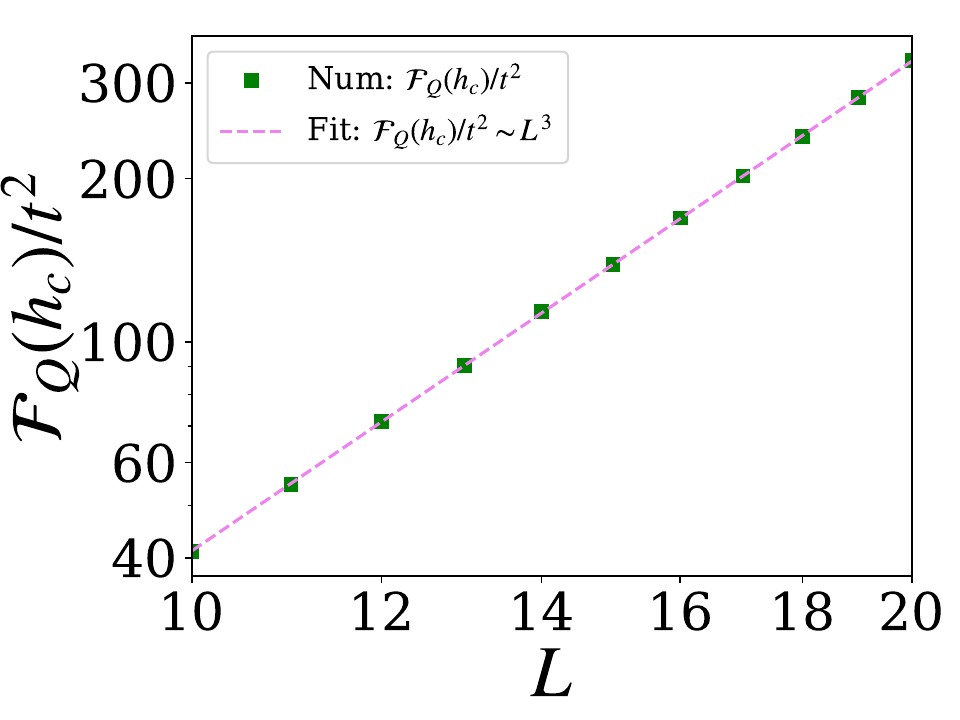}
		\put(-20,23){(c)}
		\includegraphics[width=.49\linewidth]{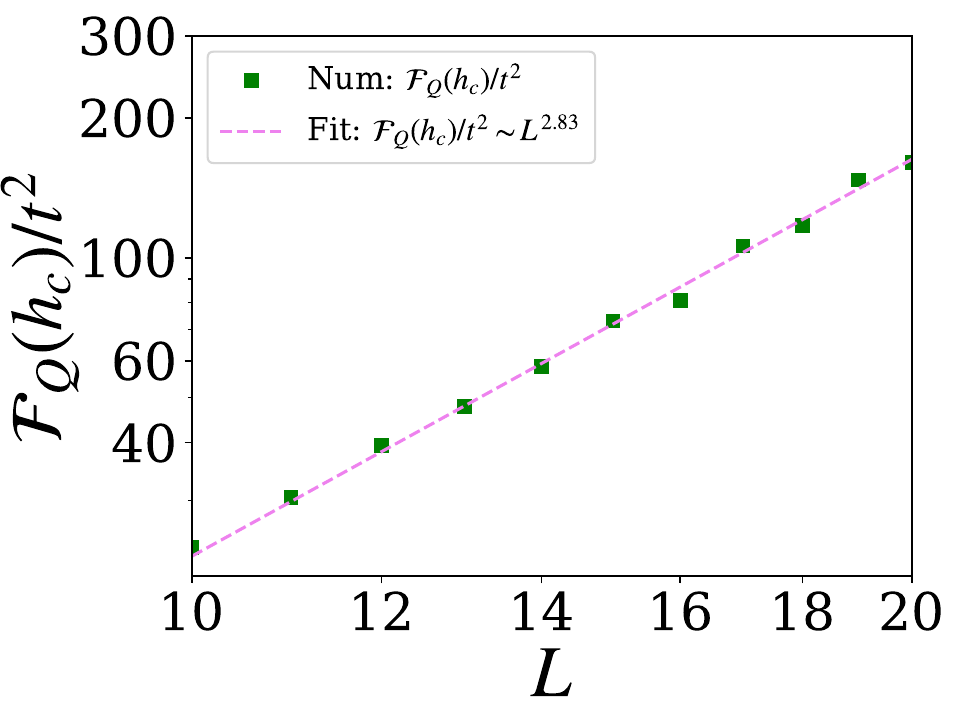}
		\put(-20,23){(d)}
		\caption{The normalized QFI for the half-filled initial Neel state as a function of $h/J$ for different system sizes for (a) $\Delta{=}0$; and (b) $\Delta{=}1$. The long-time limit of $\mathcal{F}_Q/t^2$ at the transition point $h{=}h_c$ is plotted as a function of $L$ for: (c)  $\Delta{=}0$; and (d) $\Delta{=}1$.
		} 
		\label{fig:mqfih0ND}
	\end{figure}

	We now go beyond single excitation and extend our results to systems that consist of multi-particles  which may interact with each other. We consider the XXZ Hamiltonian
	\begin{align}\label{XXZhamiltonian}
		H =& -\frac{J}{2}\sum_{l=1}^{L-1} (\sigma^x_l \sigma_{l+1}^x + \sigma^y_l \sigma_{l+1}^y + \Delta \sigma^z_l \sigma_{l+1}^z) \nonumber\\
		&+ h \sum_{l=1}^L l \sigma^+_l \sigma^-_l~,
	\end{align}
	where $\Delta$ represents the strength of the interaction between particles and $\sigma_{l}^{x,y,z}$ are the Pauli matrices acting on site $l$. 
	In the single excitation subspace and with $\Delta{=}0$, the Hamiltonian in Eq.~(\ref{XXZhamiltonian}) reduces to the single-particle case in Eq.~(\ref{Hstark}). 
	We consider the time evolution of a system initially prepared in a Neel state, i.e. $\ket{\Psi(0)}=\ket{1,0,\cdots,1,0}$ for even $L$ and $\ket{\Psi(0)}=\ket{1,0,\cdots,0,1}$ for odd $L$.  
	More details about the time evolution and complex pattern of many-body Bloch oscillations are available in Appendix \ref{B}.
	We first put our focus on $\Delta {=}0$ (non-interacting case) and $\Delta{=}1$ (isotropic Heisenberg Hamiltonian).
	The results for the more general case of $0{<}\Delta{<}1$ are presented and discussed in the next section.
	In Figs.~\ref{fig:mqfih0ND}(a) and \ref{fig:mqfih0ND}(b), we plot the long time limit of $\mathcal{F}_Q/t^2$ as a function of $h/J$ for various system sizes $L$ for $\Delta{=}0$ and $\Delta{=}1$, respectively.
	In both interacting and non-interacting regimes, the normalized QFI peaks at the transition point $h_c$ and it shows algebraic decay of the form $\mathcal{F}_Q/t^2 \sim h^{-2}$ in the localized phase, which is similar to the single excitation case.
	However, unlike the single excitation scenario, even in the localized phase the QFI  shows size dependence. 
	Furthermore, one can see that in the presence of interaction, the behavior of the QFI qualitatively changes as: (i) it shows fluctuations in the vicinity of the transition point; and (ii) its value becomes smaller than the non-interacting case (i.e. $\Delta=0$).
	To extract the scaling with system size, in Figs. \ref{fig:mqfih0ND}(c) and \ref{fig:mqfih0ND}(d) 
	we plot the long-time value of $\mathcal{F}_Q/t^2$ at $h{=}h_c$ as a function of system size $L$ for $\Delta{=}0$ and $\Delta{=}1$, respectively.
	In both cases, the scaling suggests $\beta > 2$, which indicates a quantum-enhanced scaling.
	Especially, in the non-interacting case $\Delta=0$, it takes the form of $\mathcal{F}_Q (h_c) \sim t^2 L^3$.
	The difference of $\beta > 2$ here, compared to the single-particle result of $\beta{=}2$ at the transition point, is related to the number of excitations that will be further explained in the next section.

It is worth emphasizing that the transition point $h_c$ decreases as the system size increases, see Fig.~\ref{fig:2}(a) and Figs.~\ref{fig:mqfih0ND}(a)-(b). In other words, in the thermodynamic limit the ergodic regime over which the quantum-enhanced sensitivity can be obtained shrinks to a vanishingly small region.
This has an implication for the scaling of the QFI. For any given $h$, as the system size increases, at some point the system enters the localized phase and hence the QFI becomes size independent.
For instance, in the case of the single-particle probes, for any given $h$ when the system size exceeds $L_c{\simeq} 8J/h $ the probe becomes localized and the QFI saturates. This means that quantum-enhanced sensitivity can only be observed over finite sizes, which is in agreement with the results of Ref.~\cite{Boixo2007generalized} that at best predicts Heisenberg scaling (i.e. $\mathcal{F}_Q\sim t^2L^2$) for single-body interacting systems, such as our Stark Hamiltonian. Note that although the quantum-enhanced sensitivity does not survive in the thermodynamic limit, in practice, all quantum probes are essentially finite and thus their scaling over short system sizes matters more than the true thermodynamic limit.

	\begin{figure}
		\includegraphics[width=.49\linewidth]{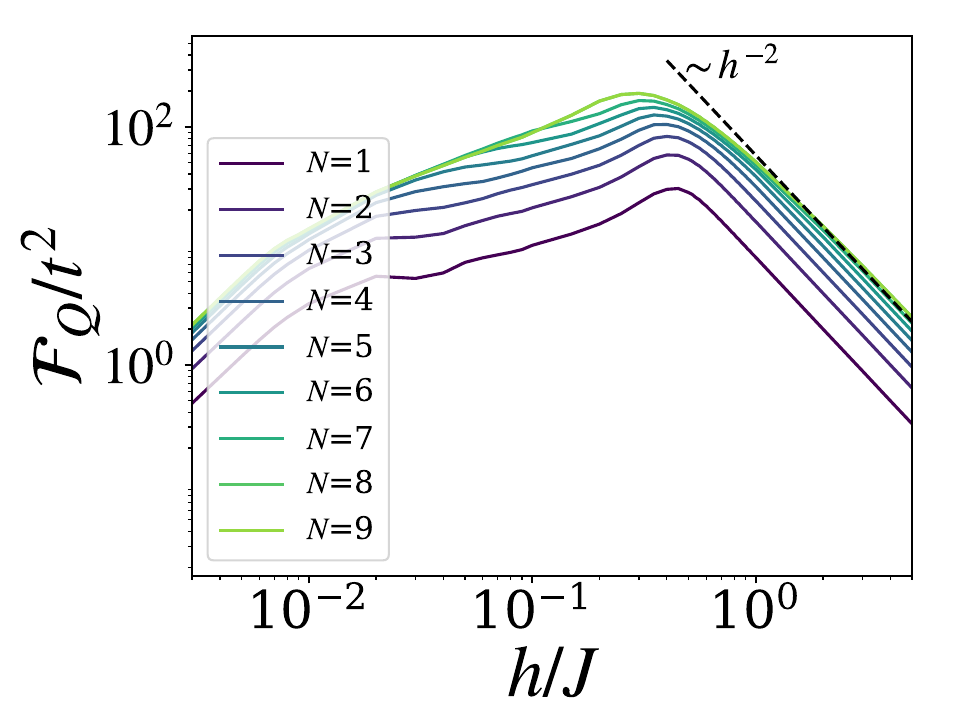}	
		\put(-90,70){(a)}	
		\includegraphics[width=.49\linewidth]{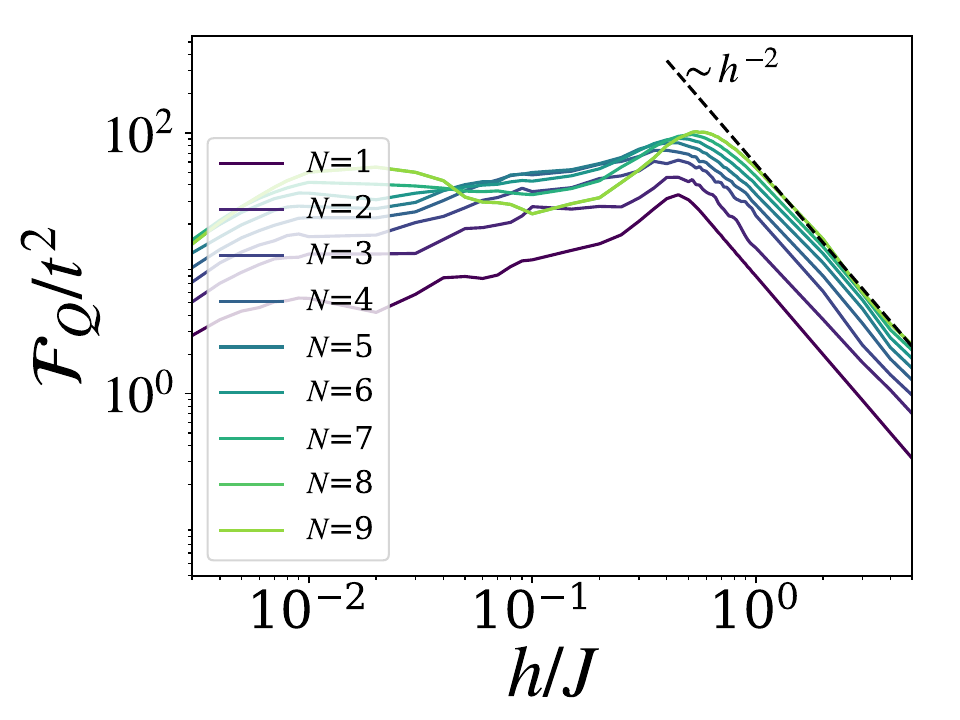}
		\put(-90,70){(b)}\\
		\includegraphics[width=.49\linewidth]{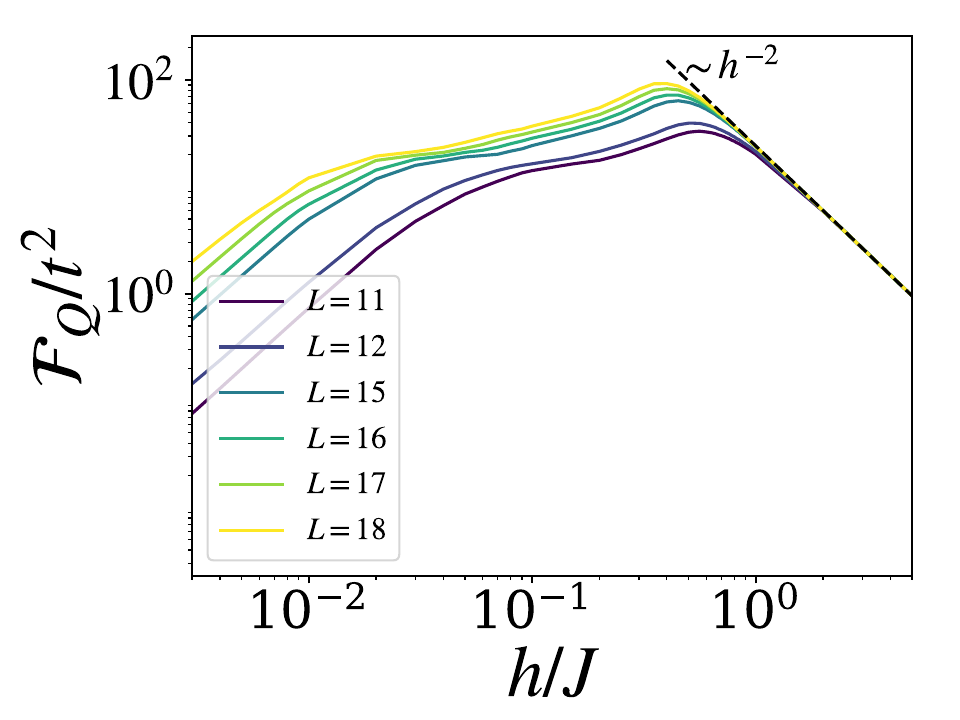}
		\put(-90,70){(c)}
		\includegraphics[width=.49\linewidth]{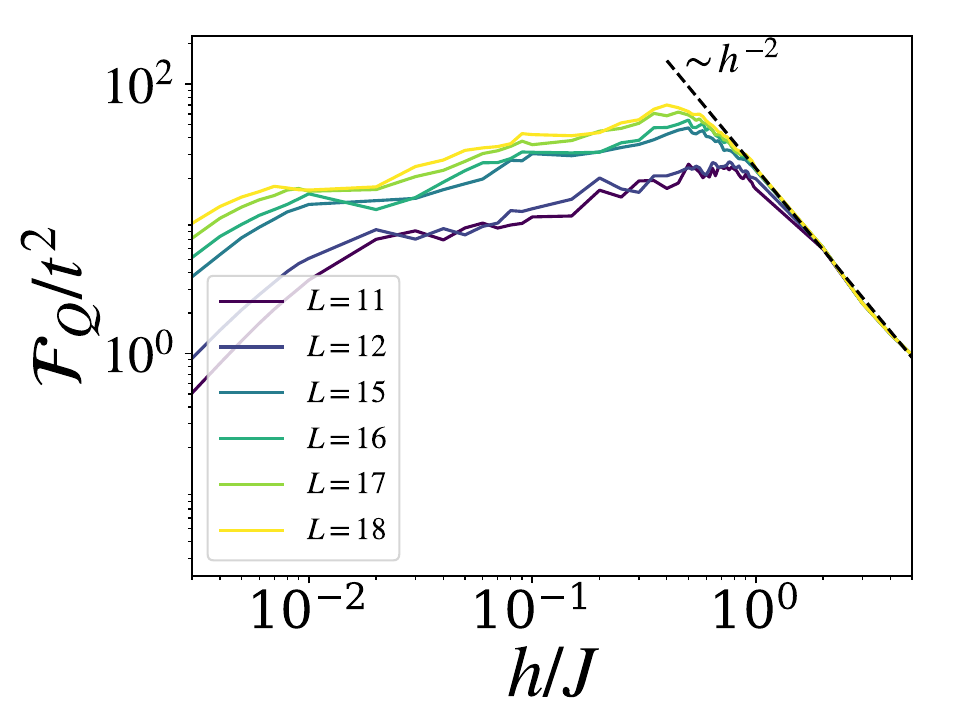}
		\put(-90,70){(d)}
		\caption{The long-time limit of $\mathcal{F}_Q/t^2$ in  system of size $L{=}17$ for initial states with different excitations as a function of $h/J$ for: (a) $\Delta{=}0$; and (b) $\Delta{=}1$. The long-time limit of $\mathcal{F}_Q/t^2$ for a system with the fixed number of excitations $N {=}3$ and various system sizes for: (c) $\Delta{=} 0$; and (d) $\Delta{=} 1$.}	 
		\label{fig:mexcqfih0ND}
	\end{figure}

	\begin{figure}[b]
		\includegraphics[width=.49\linewidth]{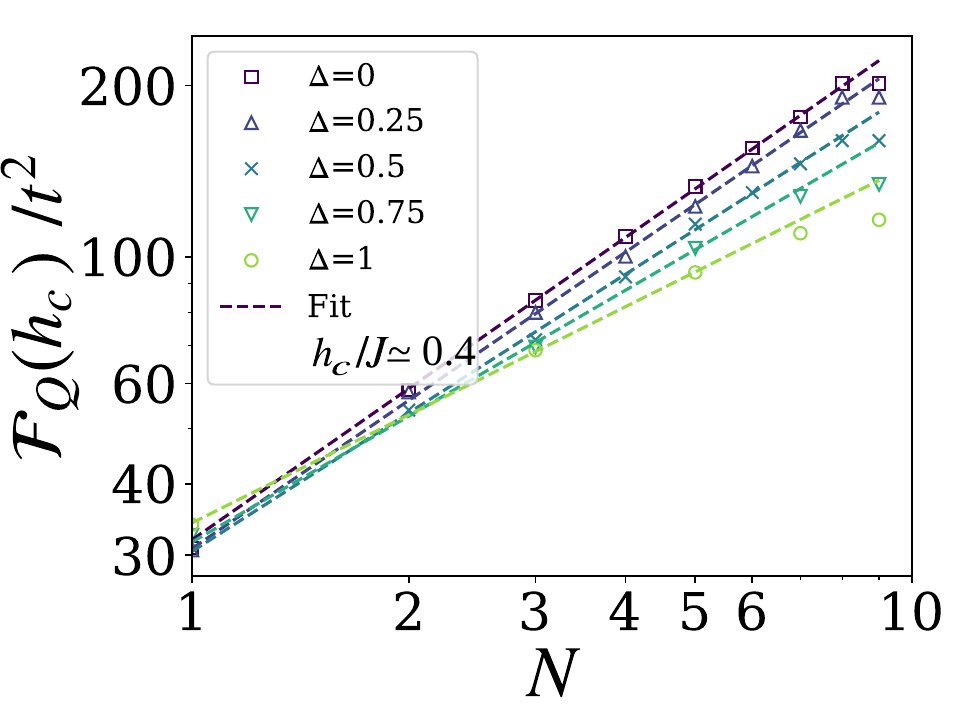}	
		\put(-95,87){(a)}
		\put(20,87){(b)}	
		\includegraphics[width=.49\linewidth]{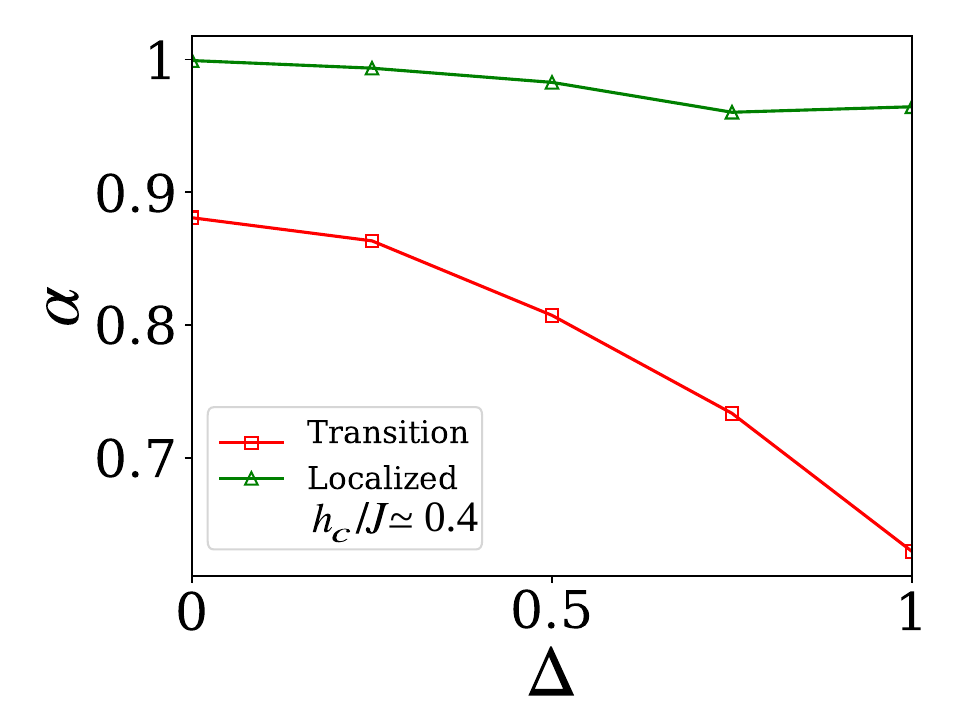}
		\caption{ (a) The scaling of long-time normalized QFI at the transition point with respect to the excitation number $N$,  in a system of size $L{=}17$ for different interaction strengths $\Delta$ which indicates $\mathcal{F}_Q(h_c)/t^2 {\sim} N^\alpha$. (b) Exponent $\alpha$ as a function of $\Delta$ for two choices of $h$ at the transition points $h_c$ and in the localized phases at $h/J{=}5$.
		}
		\label{fig:mexcNfitD}
	\end{figure}
	
\section{Effect of excitations}%
	In the previous section, we focused on the half-filling case, namely $N{=}L/2$. In this section, we investigate the impact of the number of excitations in more details. For the sake of symmetry, we consider odd system sizes $L$ and choose the initial states to be symmetric around the center of the chain. For instance, if the system size is $L{=}7$ then the initial state with $N=2$ and $N=3$ excitations are set to  $\ket{0,0,1,0,1,0,0}$ and $\ket{0,1,0,1,0,1,0}$, respectively. 
	In Figs.~\ref{fig:mexcqfih0ND}(a) and \ref{fig:mexcqfih0ND}(b) the normalized QFI is plotted for different excitation numbers $N$ in a system of size $L{=}17$ for $\Delta{=}0$ and $\Delta{=}1$, respectively. Notably, increasing the excitations always enhances the QFI, indicating better sensitivity. In addition, in the localized phase the QFI shows excitation-dependence. In order to discriminate the QFI enhancement due to the system size from the excitation number, in  Figs.~\ref{fig:mexcqfih0ND}(c) and \ref{fig:mexcqfih0ND}(d) we plot the long-time behavior of $\mathcal{F}_Q/t^2$ as a function of $h/J$ for a fixed number of excitation $N{=}3$ for various choices of $L$. The scaling of the normalized QFI at the transition points in Fig.~\ref{fig:mexcqfih0ND}(c) and (d) are discussed in Appendix~\ref{C}. 
    	These observations suggest the following ansatzes for the QFI in many-body probes with multiple excitations
	\begin{eqnarray}\label{eq:QFI_scaling_N_L}
		\mathcal{F}_{Q} & \sim h^{-2} t^2 L^0 N^\alpha,  \;\; \text{   for } h>h_c~, \cr
		\mathcal{F}_Q & \sim t^2 L^\beta  N^\alpha,  \;\;\;\;\;\;\;\; \text{   for } h \le h_c~,
	\end{eqnarray}
	where $\alpha$ is the exponent which quantifies the dependence on the excitation number.
	The size dependence completely disappears for $h>h_c$ when the system enters the localized phase. This can be understood as every excitation in the system can only fluctuates locally and thus dependence on the system size disappears. In this situation, the sensitivity is only enhanced by either waiting for longer times $t$ or increasing the number of excitations $N$. The reason behind this is that every excitation localizes at some spatial part of the probe where it fluctuates locally. Every small changes from each of these excitations contributes in enhancing the QFI of the system.

    In Fig.~\ref{fig:mexcNfitD}(a) for a fixed system of size $L{=}17$, we plot the long-time value of $\mathcal{F}_Q/t^2$ as a function of $N$ which clearly shows an algebraic growth. To complete the analysis, in Fig.~\ref{fig:mexcNfitD}(b) the exponent $\alpha$ is depicted as a function of  $\Delta$, for two choices of $h$, one in the localized phase, i.e. $h/J{=}5$, and one at the transition point $h{=}h_c$. As the figure shows, while increasing the excitations always enhances the QFI, its rate of enhancement decreases by increasing the interaction strength $\Delta$.
	This is an interesting observation which can be related to various configurations that the system gets due to the interaction term $\Delta$ and effectively enhances the localization~\cite{yousefjani2023mobility}. 
	Finally, it is worth emphasizing that in the case of half-filling, namely $N{=}L/2$, and in the range of $h\le h_c$ (i.e. extended phase) Eq.~(\ref{eq:QFI_scaling_N_L}) suggests that the QFI scales as $\mathcal{F}_Q \sim t^2 L^{\beta+\alpha}$. 
	The main advantage of Eq.~(\ref{eq:QFI_scaling_N_L}) is to separate the dependence on $L$ and $N$. Therefore, one can extract $\beta$ from the single excitation analysis, shown in Fig.~\ref{fig:2}(c), and the exponent $\alpha$ can be read from Fig.~\ref{fig:mexcNfitD}(b). 
	For instance, at the transition point $h{=}h_c$ the single excitation analysis gives $\beta{=}2$ and from Fig.~\ref{fig:mexcNfitD}(b) one reads $\alpha{=}0.88$ at $\Delta{=}0$ and $\alpha=0.63$ at $\Delta{=}1$. These lead to  $\mathcal{F}_Q\sim t^2 L^{2.88}$  and $\mathcal{F}_Q\sim t^2 L^{2.63}$ for $\Delta{=}0$ and $\Delta{=}1$, respectively. These scaling functions are very close to the fittings directly extracted in Fig.~\ref{fig:mqfih0ND}. 
 
\section{Source of the quantum-enhanced sensitivity in the Stark probe}

While the above analysis relies solely on numerical simulation, one can also shed light on the nature of quantum-enhanced sensitivity by focusing on the analytical bounds. In particular, one can decompose the Hamiltonians of Eq.~\eqref{Hstark}, for a single excitation probe, and Eq.~\eqref{XXZhamiltonian}, for the many-body probe, into a general form of $H=H_1+hH_2$, where $H_1$ accounts for the hopping term and $H_2$ describes the gradient field.  For a given initial state $\ket{\Psi(0)}$ which is independent of $h$, the time evolution is given by $\ket{\Psi_h(t)}=e^{-iHt}\ket{\psi(0)}$. In Ref.~\cite{Boixo2007generalized}, it has been shown that the QFI of the quantum state $\ket{\Psi_h(t)}$  is fundamentally bounded by $\mathcal{F}_Q {\le} t^2 ||H_2||^2$, where $||H_2||{=}\lambda_{\rm max}{-}\lambda_{\rm min}$ is the semi-norm with $\lambda_{\rm max}$ ($\lambda_{\rm min}$) being the maximum (minimum) eigenvalue of $H_2$.  Although, the term $t^2 ||H_2||^2$ is only a bound for $\mathcal{F}_Q$, it provides several insights for the scaling of the QFI. First, it clearly demonstrates the quadratic scaling with respect to time. Second, it shows that the scaling with respect to system size $L$ is connected to the semi-norm of $H_2$.

In the case of a single excitation, see  Eq.~\eqref{Hstark}, the semi-norm $||H_2||{=}(L-1)$, resulting in the bound of $\mathcal{F}_Q {\le} t^2 (L-1)^2$. This fundamental bound shows that the QFI scaling exponent can only be  $\beta{\le}2$. Note that this is not inconsistent with Fig.~\ref{fig:2}(c) in which the extended phase supports $\beta{>}2$. In fact, as mentioned before, this region is only exist in finite system sizes as in the thermodynamic limit, where the critical point $h_c{\rightarrow}0$, the region shrinks to an infinitesimal  region.

In the case of  many-body probes with $N$ excitations, the semi-norm becomes $||H_2||{=}NL{-}N^2$. For the case of half filling where $N{=}L/2$, one gets $||H_2||{=}L^2{/}4$. This implies that $\mathcal{F}_Q {\le} t^2 L^4/16$. Indeed, our analysis in section~\ref{MBp} shows that $\mathcal{F}_Q {\sim} t^2 L^{\beta+\alpha}$, in which $\beta{\le}2$ comes from the single excitation analysis and $\alpha{\le}1$ is extracted from Fig.~\ref{fig:mexcNfitD}(b). This means that the QFI scales as $\mathcal{F}_Q{\sim}t^2 L^3$,  which is clearly lower than the prediction of the analytic bound given by $t^2||H_2||^2{\sim}t^2L^4$.

\section{Impact of decoherence}	
\begin{figure}[t]
	\includegraphics[width=.49\linewidth]{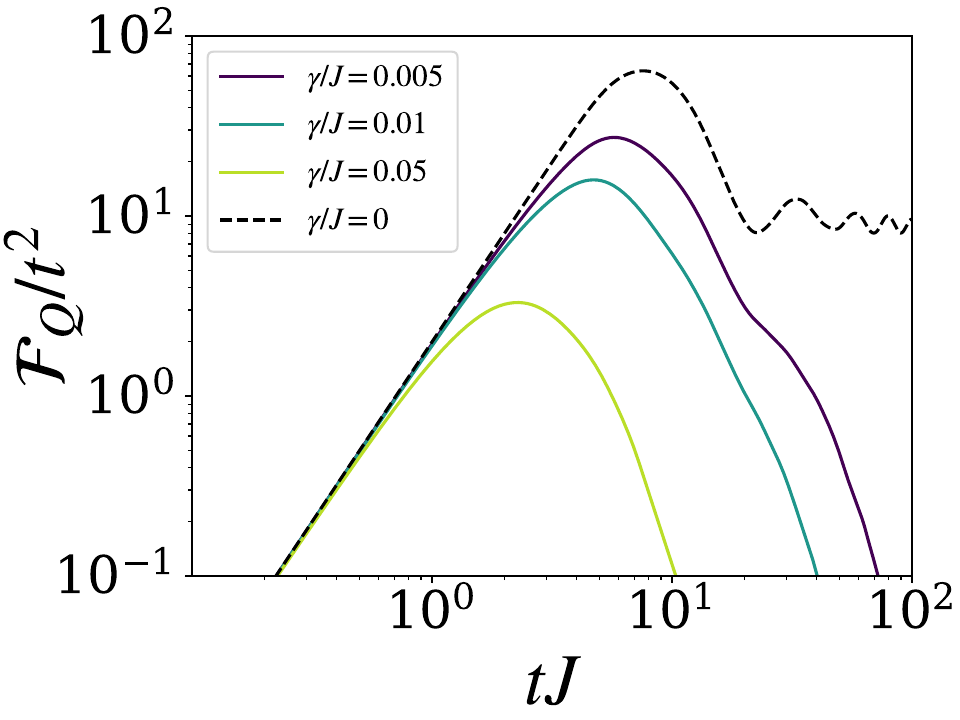}	
	\put(-90,22){(a)}
	\includegraphics[width=.49\linewidth]{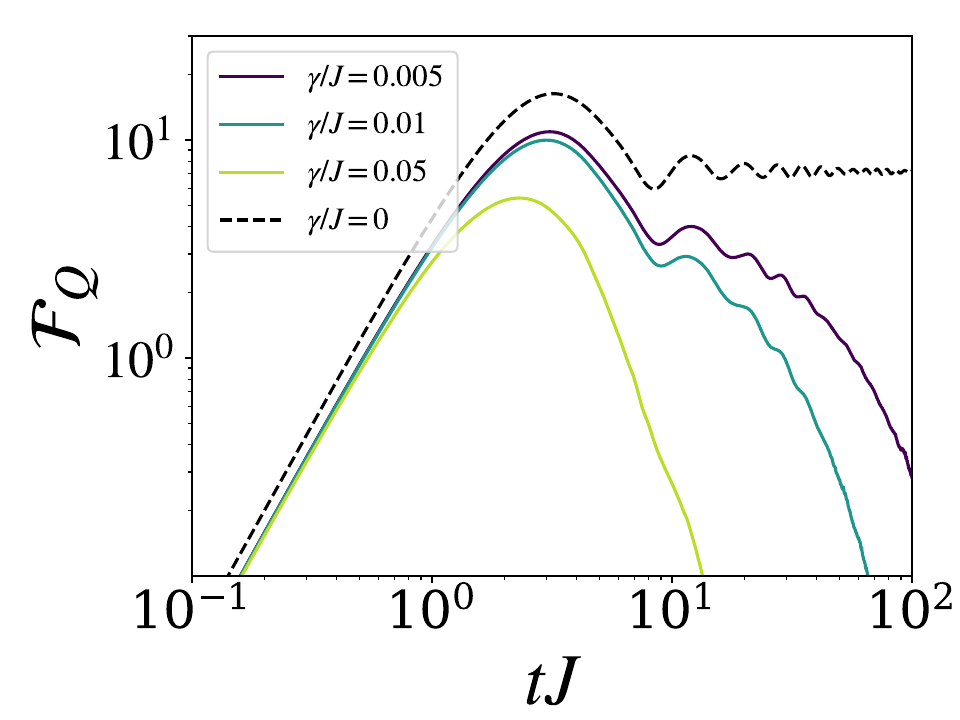}	
	\put(-90,22){(b)}\\	
	\includegraphics[width=.49\linewidth]{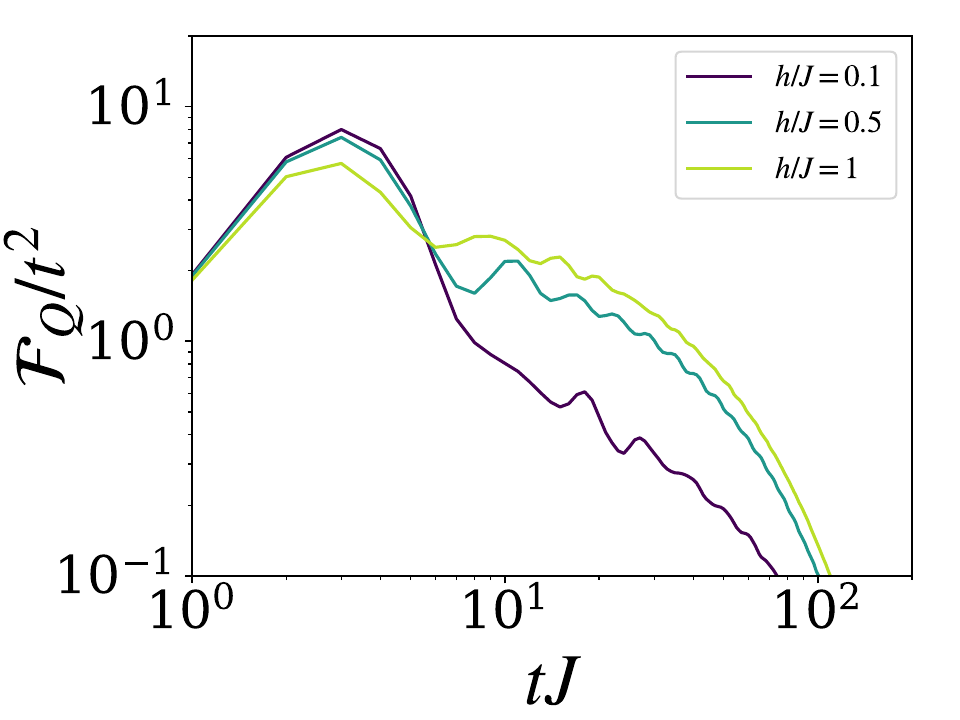}	
	\put(-90,22){(c)}
	\includegraphics[width=.49\linewidth]{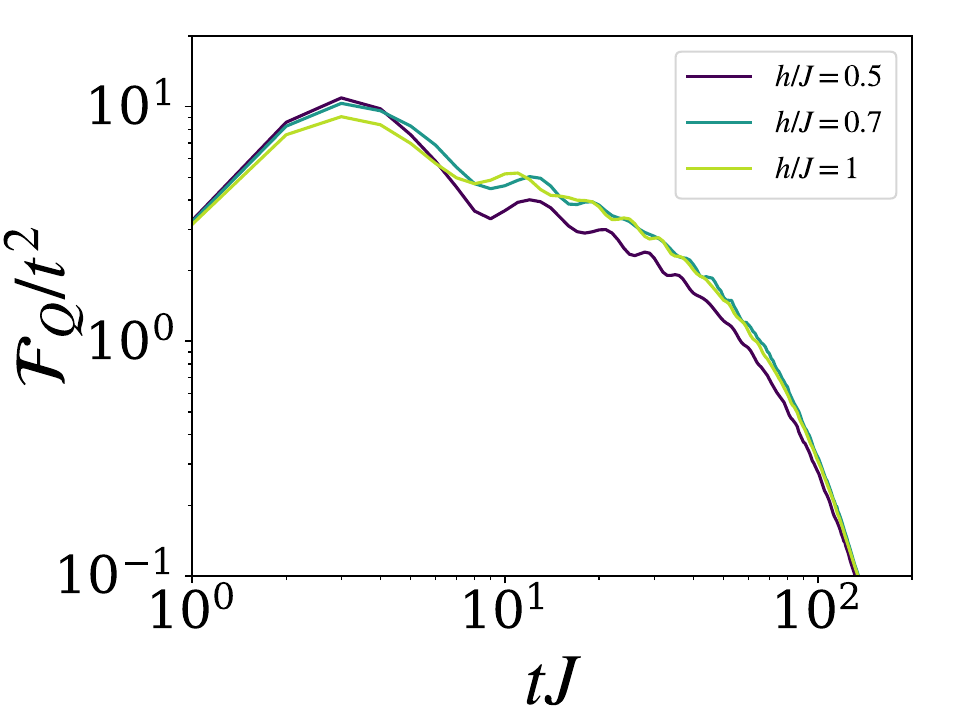}
	\put(-90,22){(d)}	
	\caption{  
	The time evolution of $\mathcal{F}_Q/t^2$ for different values of $\gamma$ is plotted for: (a) the single-excitation probe with the system size $L=16$ at $h/J{=}0.1$; and (b) the half-filled excitation probe with the system size $L{=}6$ for $\Delta=0$ at $h/J{=}0.5$.		
	The time evolution of $\mathcal{F}_Q/t^2$ for dephasing strength $\gamma/J{=} 0.005$ and different values of $h/J$  is shown as a function of time for: (c) the single-excitation probe with the system size $L{=}16$; and (d) the half-filled excitation probe with the system size $L{=}6$ with $\Delta=0$.}
	
	\label{fig:Decoherence}
\end{figure}

In practice, quantum systems are not isolated, and thus their dynamics cannot be described by unitary evolution. In fact, interaction with the environment can significantly affect the quality of quantum probes \cite{Escher_2011,Demkowicz_Dobrza_ski_2012}. To investigate this, we consider our quantum probe under the effect of dephasing. The dynamics is given by the Lindblad master equation
\begin{equation}
	\dot{\rho}= -i [H,\rho]+\gamma \sum_{l=1}^{L}  \l(\mathcal{L}_l \rho \mathcal{L}^\dagger_l - \frac{1}{2} \{\mathcal{L}^\dagger_l\mathcal{L}_l ,\rho\}\r)~,
\end{equation}
where $\rho$ is the density matrix of the probe, $\gamma$ is the dephasing rate, and $\mathcal{L}_l$ represents the Lindblad operators of site $l$. For pure dephasing, in general, the Lindblad operators are given by $\mathcal{L}_l{=} \sigma^z_{l}$. In the single-excitation probe,  the Lindblad operators can also be written as $\mathcal{L}_l{=} \mathbb{I} -2 \ket{l}\bra{l}$, in which $\mathbb{I}{=}\sum_{l=1}^{L}\ket{l}\bra{l}$. 
In Fig.~\ref{fig:Decoherence}(a), we plot the normalized QFI, namely $\mathcal{F}_Q/t^2$, at $h/J{=} 0.1$ as a function of time for different values of dephasing rates $\gamma$ for the single-particle probe of size $L=16$. As the figure shows, by increasing $\gamma$ the QFI ultimately decays to zero and cannot grow quadratically in time. Similarly, for the multi-particle probe of size $L{=}6$ with $N{=}3$ excitations (i.e. half filling), we plot the normalized QFI for $h/J{=}0.5$ as a function of time for various choices of $\gamma$ in Fig.~\ref{fig:Decoherence}(b). The results show similar behavior with the single-excitation probe as the normalized QFI eventually goes to zero as time increases. This is because the steady state of the probe under dephasing dynamics is the featureless maximally mixed state which carries no information about the parameters of the system. 
One can also fix the dephasing rate and investigate the evolution of the QFI for different choices of $h$. In
Figs.~\ref{fig:Decoherence}(c)-(d), we present the normalized QFI as a function of time for $\gamma/J= 0.005$ and various choices of $h/J$ for the single- and multi-particle probes, respectively. 
Interestingly, by increasing $h$ the probe gets more localized and thus the action of dephasing becomes less destructive as the wave function is made of fewer components in its superposition.  

\section{Practical implementation}%

\begin{figure}[b]
	\includegraphics[width=.49\linewidth]{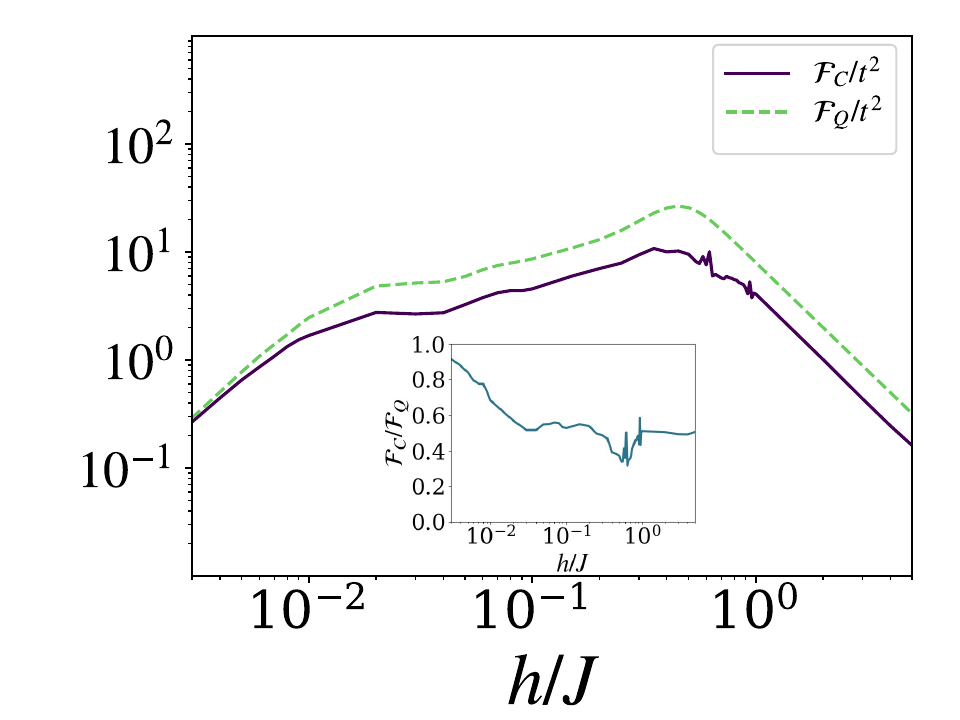}	
	\put(-90,70){(a)}
	\includegraphics[width=.49\linewidth]{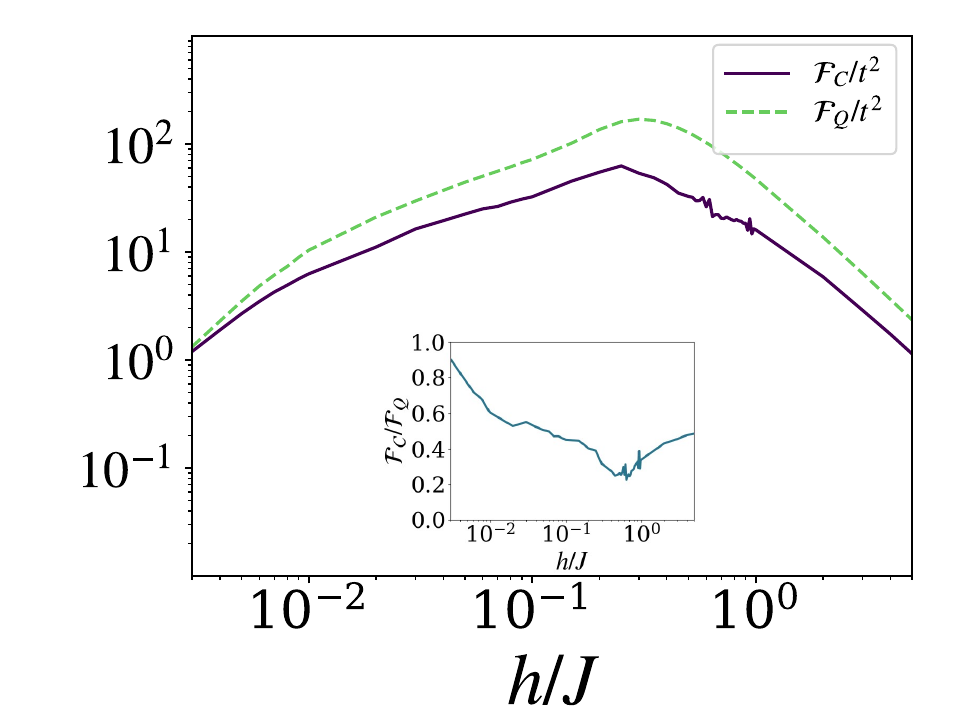}
	\put(-90,70){(b)}	
	\caption{The long-time limit of $\mathcal{F}_{C}$ (purple solid lines) and $\mathcal{F}_{Q}$ (green dashed lines) as a function of $h/J$ for the system size $L{=}16$ are plotted for the initial state of: (a) a single excitation state $\ket{1}$; and (b) the half-filled Neel state with  $\Delta{=}0$. The insets, depicting the ratio of $\mathcal{F}_C/\mathcal{F}_Q$ versus $h/J$, show that our specific choice of simple measurements can reach around $50\%$ of the ultimate precision bound, determined by the QFI.   
	}
	\label{fig:CFI}
\end{figure}

As discussed before, QFI determines a lower bound on the precision of an unbiased estimator given optimal measurements. In practice, however,  performing such an optimal measurement might be very challenging. In this section, we consider simple available measurements in both single- and multi-particle cases. For the single excitation sensors, we consider a simple position measurement described by projectors $\{ \pi_l{=}\ket{l}\bra{l}\}$, where $l{=}1,2,\cdots,L$. Therefore, the probability of each outcome is $p_l(t){=} |\braket{l|\Psi(t)}|^2$. The obtainable  precision by this measurement is bounded by the CFI, see Eq.~(\ref{CFI}). 
In Fig.~\ref{fig:CFI}(a), we compare the long-time behavior of the CFI and the QFI in a single-excitation probe of size $L{=}16$ as a function of $h/J$. As the figure shows, the CFI qualitatively follows the QFI, though it always takes lower values. In the inset, we plot the ratio $\mathcal{F}_C/\mathcal{F}_Q$ as a function of $h/J$ which shows that our simple measurement is almost optimal in the extended phase (i.e. $\mathcal{F}_C/\mathcal{F}_Q\simeq 1$) and it becomes sub-optimal (i.e. $\mathcal{F}_C/\mathcal{F}_Q\sim 0.5$) as localization strength is enhanced. Interestingly, the lowest ratio of $\mathcal{F}_C/\mathcal{F}_Q$ takes place around the transition point showing that our simple measurement scheme cannot efficiently extract information from the complex critical quantum state.

In the case of a multi-particle probe, we measure the configuration of particles which is obtained by a simple particle detection imaging. The measurement can be described by $\{ \pi_{_{\mathbf{Z}}}{=}\ket{\mathbf{z}} \bra{\mathbf{z}}\}$, where $\ket{\mathbf{z}}$ accounts for all possible particle configurations in the probe. For instance, in a probe of size $L$ with $N$ particles (i.e. excitations), there are ${L\choose N}$ different configurations. In the case of $N{=}1$, the problem reduces to the position measurement, discussed above.
Each configuration is thus observed by the probability
\begin{equation}\label{eq:CFI_Probability_multi}
    p_{_{\mathbf{z}}}(t){=} |\braket{\mathbf{z}|\Psi(t)}|^2,
\end{equation}
for which one can obtain the corresponding the CFI.
In Fig.~\ref{fig:CFI}(b), we compare the long-time behavior of the CFI and the QFI in a half-filled probe of size $L{=}16$ (with $N{=}L/2$ excitations) initialized in the Neel state as a function of $h/J$. Similar to the case of single excitation, the CFI qualitatively follows the QFI, though with lower values. 
In the inset, we plot $\mathcal{F}_C/\mathcal{F}_Q$ as a function of $h/J$. Similar to the single excitation case, it shows that a simple configuration measurement is almost optimal in the extended phase and becomes sub-optimal with $\mathcal{F}_C/\mathcal{F}_Q\simeq 0.5$  as $h/J$ increases. Again the lowest ratio of $\mathcal{F}_C/\mathcal{F}_Q$ is obtained around the transition point where our simple measurement scheme cannot fully extract the information from the complex critical quantum state of the system.

\section{Maximum Likelihood  Estimation}
The Cram\'{e}r-Rao inequality only indicates the lower bound of the precision of the probe using an optimal estimation algorithm with optimal measurements. In order to see the real performance of our probe, we specify a measurement basis and use an estimator. 
Maximum likelihood estimation (MLE) algorithm is recognized as an optimal estimation algorithm for a large measurement samples. For a given measured data set $X$, the estimation for $h$, denoted $h_{es}$, that maximizes the likelihood can be written in the form of  
\begin{equation}
h_{es}{=}\mathrm{arg}~\underset{\{h\}}{\mathrm{max}}P(X|h)~,
\end{equation}
where $P(X|h)$ is the likelihood. 

On the other hand, in Bayesian estimation the estimated value is obtained at the point where the posterior probability distribution $P(h|X){=} P(X|h)P(h)/P(X)$ is maximal, where $P(h)$ is the prior probability distribution, and $P(X)$ is the normalization factor.
However, when the prior probability distribution is uniform and the sample size is sufficiently large, both estimators converge to identical results \cite{berger2013statistical,lehmann2006theory}~. This is because, in the limit of a large sample size, the central limit theorem ensures that the likelihood function approaches a Gaussian distribution. When the sample size is sufficiently large, this Gaussian distribution becomes sharply peaked, concentrating over a narrow interval within the support of the prior. In this paper, we rely on maximum likelihood estimation and thus the choice of prior becomes irrelevant.  

In order to obtain our data set, one needs to specify a measurement basis which we take it to be the position measurement whose outcomes appear with probability $p_{_{\mathbf{z}}}(t)$ given in Eq.~(\ref{eq:CFI_Probability_multi}). By repeating the measurement for $\mathcal{M}$ times, each outcome $|\mathbf{z}\rangle$ appears for $n_{_{\mathbf{z}}}$ times, such that $\sum_{_{\mathbf{z}}} n_{_{\mathbf{z}}}{=}\mathcal{M}$. Therefore, the likelihood can be written as
\begin{equation}\label{Bayesian}
	P(X|h)= \begin{pmatrix}
	\mathcal{M} \\ n_{\mathbf{0}},n_{\mathbf{1}}, ... , 
	\end{pmatrix} \prod_{\mathbf{z}} p_{\mathbf{z}}^{n_{\mathbf{z}}}~.
\end{equation}

    \begin{figure}
    	\includegraphics[width=.49\linewidth]{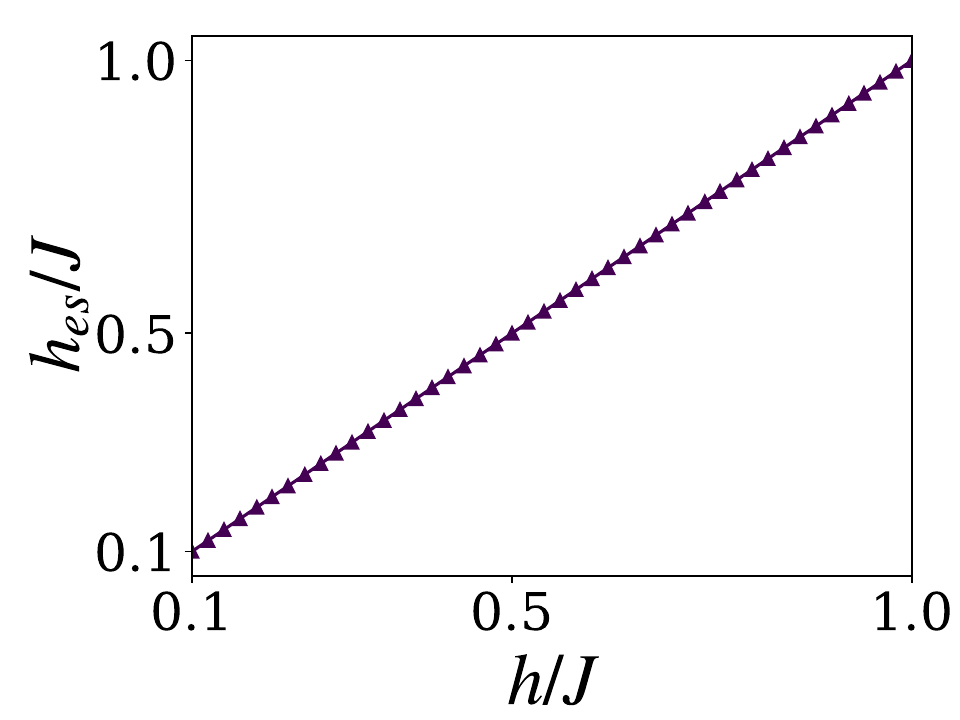}	
    	\put(-20,22){(a)}
    	\includegraphics[width=.49\linewidth]{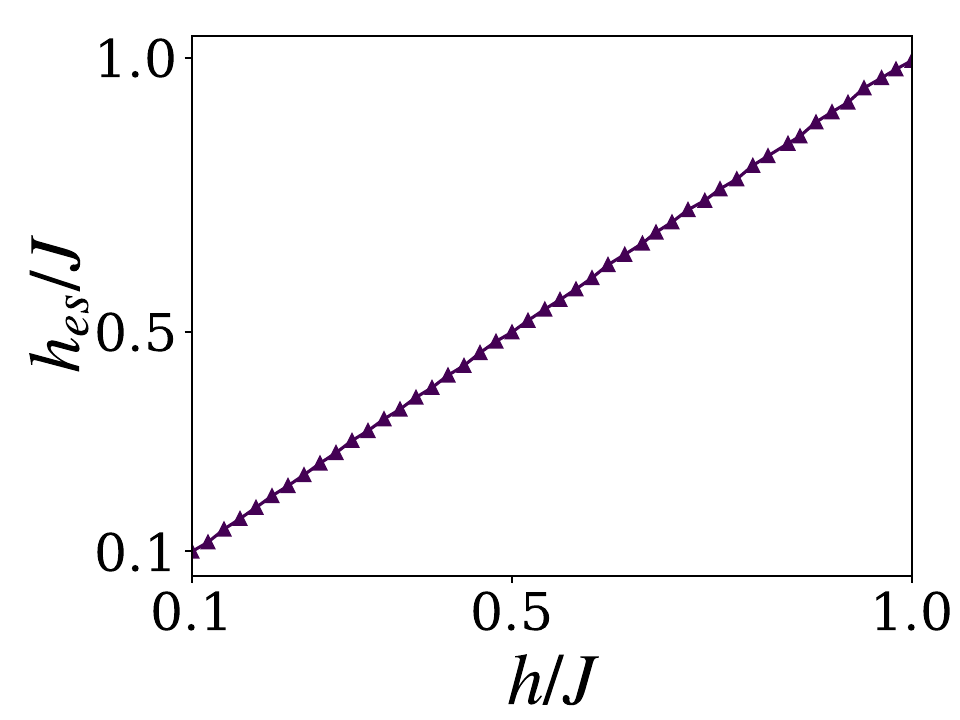}	
    	\put(-20,22){(b)}\\	
    	\includegraphics[width=.49\linewidth]{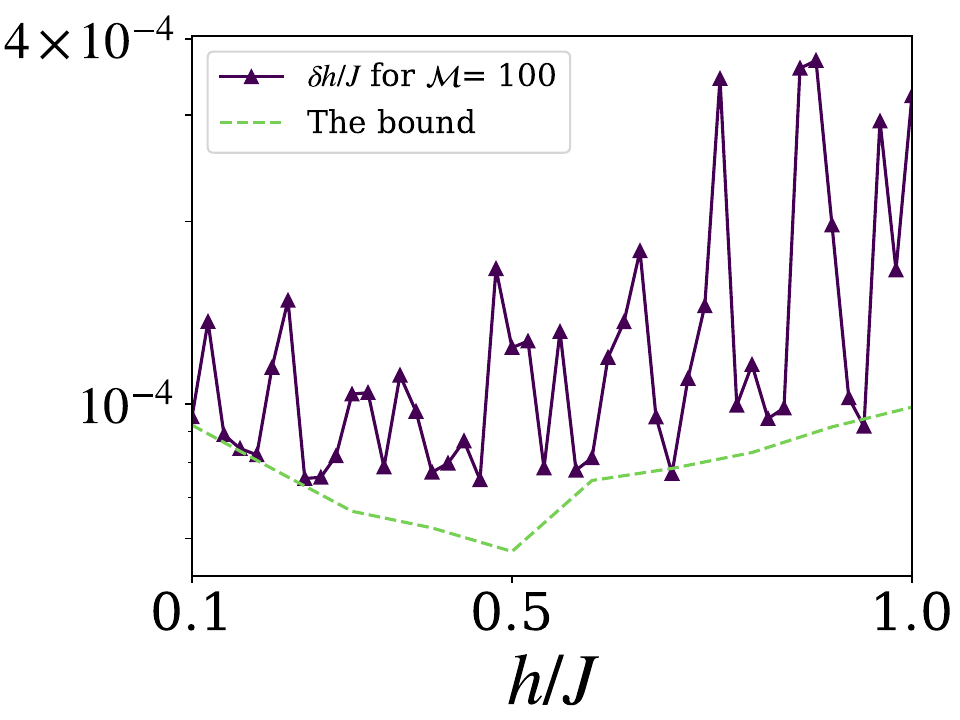}	
    	\put(-20,22){(c)}
    	\includegraphics[width=.49\linewidth]{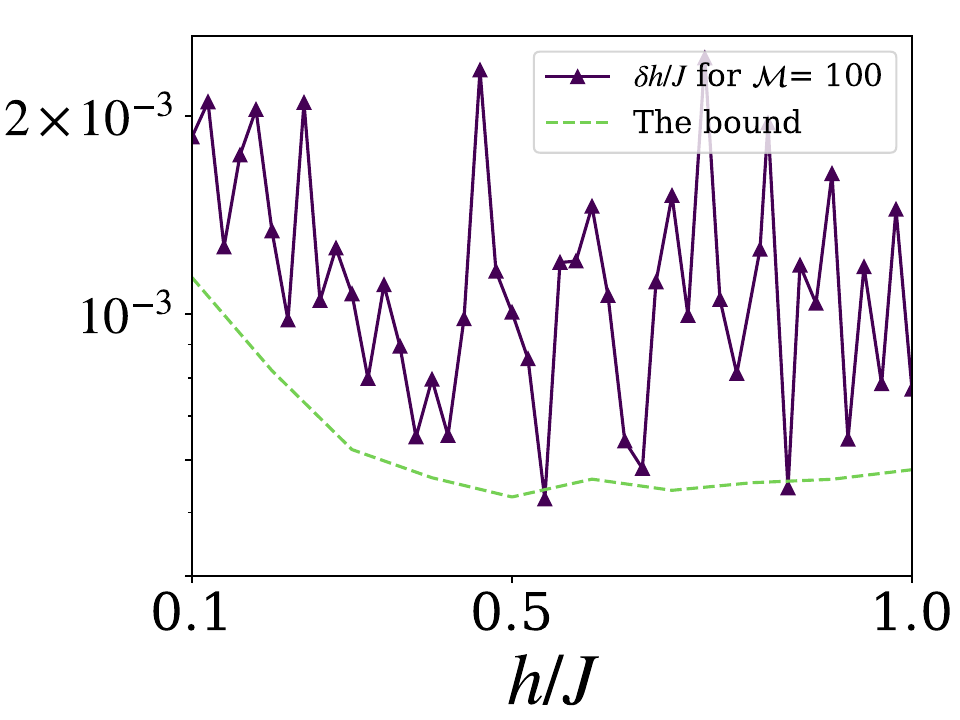}
    	\put(-90,22){(d)}	
    	\caption{ The estimated gradient field $h_{es}/J$ using the maximum likelihood estimator as a function of its true value for sample sizes of $\mathcal{M}=100$ is shown for: (a) the single-particle probe with the system size $L{=}16$; and (b) the multi-particle probe of  size $L{=}6$ and $\Delta{=}0$, initialized with the half-filled Neel state.
    	The corresponding standard deviations $\delta h$ are plotted for: (c) single-particle probe, and (d) multi-particle probe. The associated $\delta h$ of the long-time CFI, $1/\sqrt{\mathcal{M}\mathcal{F}_{C}}$, (dashed lines) are depicted for both probes.
    	}
    	\label{fig:Bayesian}
    \end{figure}

We generate random samples based on the true probability distribution of a given $h$. Using the MLE algorithm, $h_{es}$ is taken to be the point at which the likelihood takes its maximum. The uncertainty of the estimation can be quantified by the standard deviation $\delta h$ which is computed over the results of several repetitions of the procedure.  In Figs.~\ref{fig:Bayesian}(a)-(b) we show $h_{es}/J$ as a function of real $h/J$ for $\mathcal{M}{=}100$ samples in a single-excitation probe of $L{=}16$ and a multi-particle probe of $L{=}6$ and $N{=}3$ (i.e. half-filling), respectively. As the figures show, the estimation procedure works very well in both single- and multi-particle probes. In Figs.~\ref{fig:Bayesian}(c)-(d), we depict the corresponding standard deviation $\delta h/J$ for the sample size $\mathcal{M}{=}100$ as a function of $h/J$ for the single and the multi-particle probes, respectively. For comparison, we also plot the Cram\'{e}r-Rao bound $1/\sqrt{\mathcal{M}\mathcal{F}_C}$ which shows that the choice of the maximum likelihood estimator can closely converge with the theoretical bound for samples as small as $\mathcal{M}=100$. As shown in Figs.~\ref{fig:Bayesian}(c)-(d), even for $\mathcal{M}{=}100$, the uncertainty closely approaches the theoretical Cram\'er-Rao bound, indicating 
near-asymptotic performance and near-saturation of the bound. By further increasing $\mathcal{M}$ the uncertainty decreases and gets closer to the bound.

\section{Conclusion}%
We have studied the metrological aspects of the Bloch oscillations in single- and many-particle Stark systems which support a phase transition, across the entire spectrum, from an extended to a localized phase. Unlike conventional equilibrium criticality-based quantum sensors, our non-equilibrium probe does not require complex initialization and achieves quantum-enhanced sensitivity across the entire extended phase. In addition, our analysis provides an ansatz for the QFI in terms of time, probe size, and number of excitations. In the long-time limit, the QFI  scales quadratically with time in both extended and localized phases. In contrast, the precision with respect to the probe size transforms from quantum-enhanced scaling in the extended phase to size-independent behavior in the localized phase. This behavior looks very general as, for instance, in the case of multi-particle probes changing the Hamiltonian from non-interacting to interacting case does not change the qualitative behavior of the probe. Interestingly, the number of excitations always enhances the precision of the probe, though, the amount of enhancement depends on the interaction between excitations. We also show that a simple configuration measurement can achieve an estimation accuracy which is not far from the ultimate precision limit. Finally, we have demonstrated a practical scheme in which configuration measurement together with a  maximum likelihood estimation approach can closely saturate the  Cram\'{e}r-Rao inequality.  

\section{Codes and data availability}
The computer codes for the simulations of this paper are available at:\\
https://github.com/hassanmanshouri/The-QFI-of-Stark-model.git

The data used to plot the figures are available upon reasonable requests from the authors. 
	
\begin{acknowledgements}

The authors thank the Center for High Performance Computing at Shanghai Jiao Tong University for providing access to the Siyuan-1 cluster. The work of MA was partially supported by the Yangyang Development Foundation. AB acknowledges support from the National Natural Science Foundation of China (Grants No.
12050410253, No. 92065115, and No. 12274059), and the
Ministry of Science and Technology of China (Grant No.
QNJ2021167001L). HM acknowledges partial financial support by Iran Optics and Quantum Technologies Development Council. 
 HM would like to thank H. Bakhshian for valuable discussions on numerical analysis. This work has been supported by the Center for International Scientific Studies \& Collaborations (CISSC), "Ministry of Science, Research and Technology" of Iran.
	\end{acknowledgements}

\bibliographystyle{quantum}
\bibliography{reference-Critical}

\clearpage

\onecolumn
\appendix

\section{Appendix A: Single particle QFI scaling}\label{A}

\begin{figure}[h]	
	\includegraphics[width=.49\linewidth]{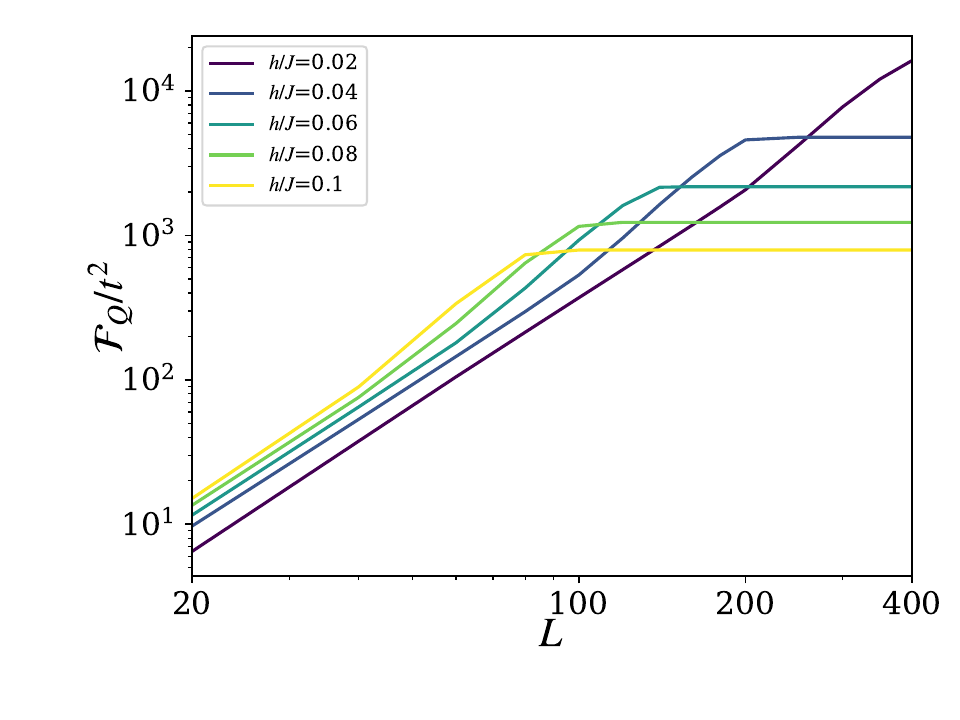}
	\caption{ The normalized QFI for different gradient fields $h$ is plotted as a function of system size $L$. The plateau for a specific $h$ resembles the localized phase in Fig.~\ref{fig:2}(a) in the main text. 
	}\label{fig:fitQFI}
\end{figure}

Further details in studying the size-dependence of the QFI that have brought us to Eqs.~\eqref{eq:QFI_Loc} in the main text are presented in Fig.~\ref{fig:fitQFI} where we plot the long-time normalized QFI as a function of system size $L$ for different choices of $h$.
Indeed, for any given $h$, one observes a power law scaling growth of $\mathcal{F}_Q/t^2$ with respect to $L$ which experiences an enhancement before turning into a plateau around the size $L {\simeq} 8J/h_c$.
Since the transition point depends on the system size the plateau sets on at different $L$ values.
One can clearly identify an improved scaling with size close to the transition point.
\\
\\

\section{Appendix B: Many-Body Bloch oscillation}\label{B}
\begin{figure}[h]
	\includegraphics[width=.32\linewidth]{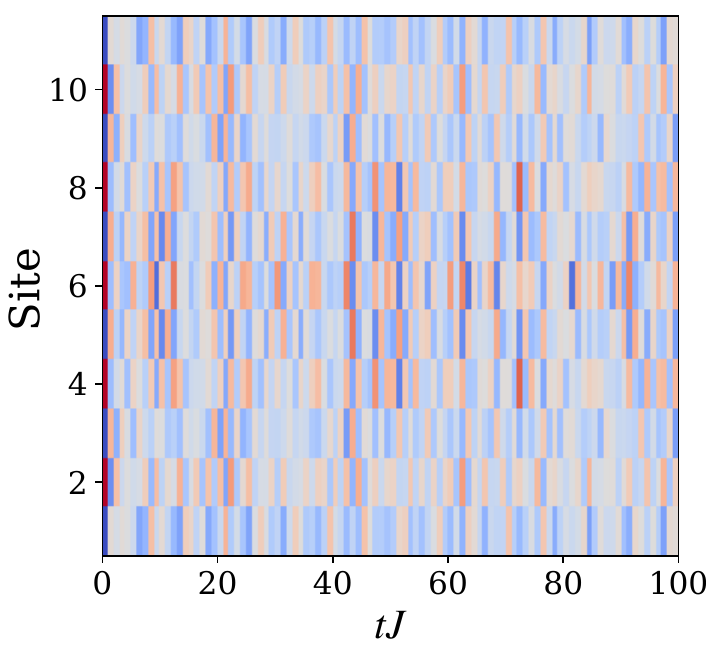}
	\put(-135,145){(a)}
	\includegraphics[width=.295\linewidth]{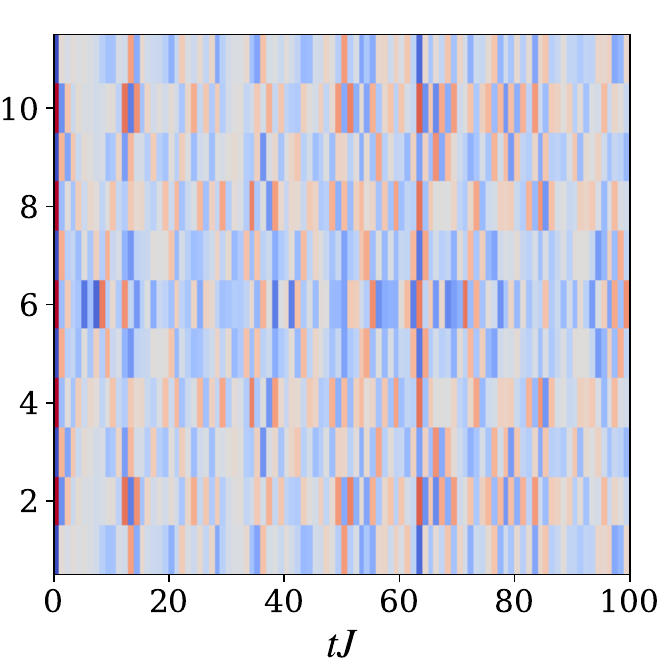}
	\put(-135,145){(b)}
	\put(10,145){(c)}
	\includegraphics[width=.36\linewidth]{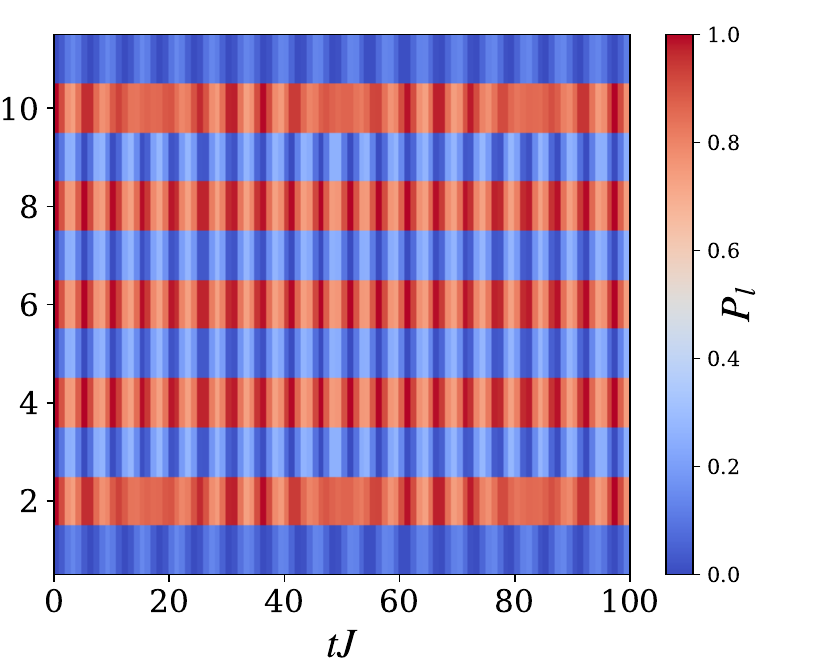}	
	\includegraphics[width=.32\linewidth]{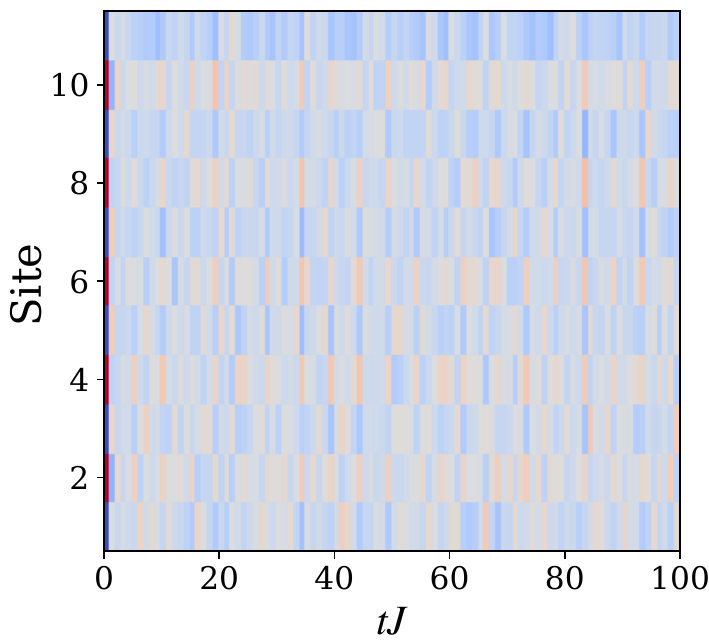}
	\put(-135,145){(d)}
	\includegraphics[width=.295\linewidth]{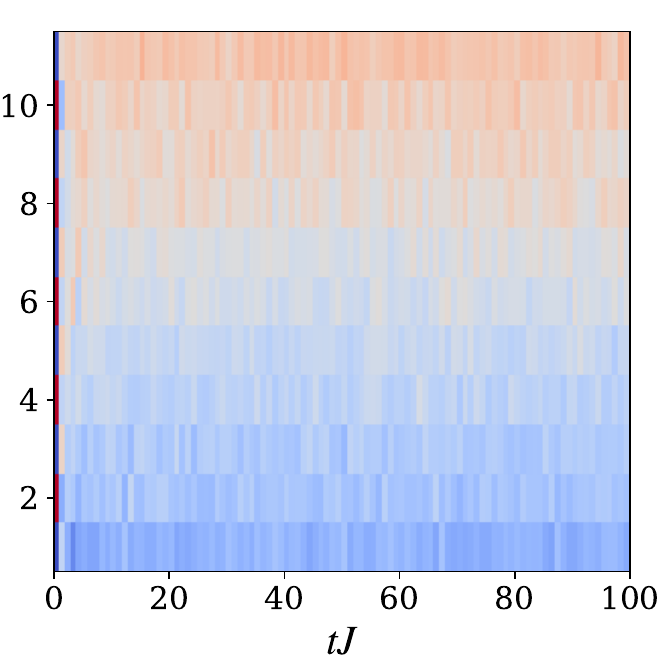}
	\put(-135,145){(e)}
	\put(10,145){(f)}
	\includegraphics[width=.36\linewidth]{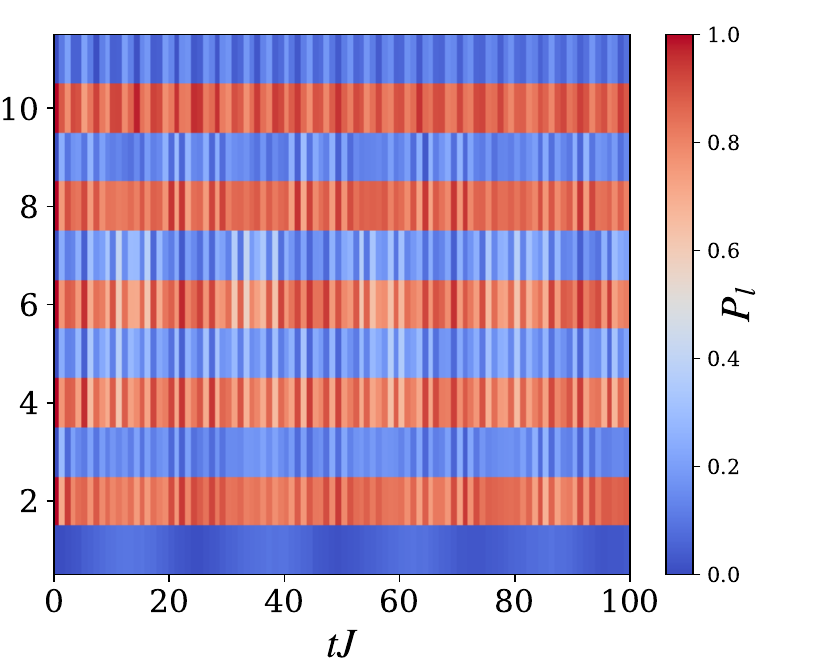}
	\caption{The probability distribution of each site $P_l$ for the initial state $\ket{\Psi_0}= \ket{01010101010}$ are depicted for the system size $L=11$ for $\Delta=0$ at (a) $h/J=0.05$ , (b) $h/J=0.4$ and (c) $h/J=5$, in extended phase, at the transition point, and in the localized phase, respectively. In (d)-(f) $P_l$ are shown for the Heisenberg Hamiltonian $\Delta=1$ at $h/J=0.05$ in the extended phase, $h/J=0.9$ at the transition point, and $h/J=5$ in the localized phase, respectively}\label{fig:mb3dp0}.
\end{figure}

In this section, we present the results of the Bloch oscillations in the many-body case to compare with the single excitation in Fig.~\ref{fig:3dp0} in the main text.
To see the behavior of the many-body excitation oscillations at different gradient fields, we compute the dynamics of the probability of having an excitation in every site of the system $P_l(t)$.
The results are presented in Fig.~\ref{fig:mb3dp0} for the system size $L=11$ with the initial state $\ket{\Psi_0}= \ket{01010101010}$.
Figs.~\ref{fig:mb3dp0}(a)-(c) are plotted for the non-interacting sites ($\Delta= 0$) in the extended phase, at the transition point and in the localized phase, respectively.
In particular, in Figs.~\ref{fig:mb3dp0}(a) the extended phase at $h/J=0.05$ shows only partial excitation revivals for a few sites at the middle of the lattice. 
However, as in Fig.~\ref{fig:mb3dp0}(b) which shows the probabilities at the transition point $h_c/J=0.4$ (extracted from Fig.~\ref{fig:mqfih0ND}(a) in the main text) complete revival for all sites are visible at particular times, e.g. at $tJ=65$.
Meanwhile, Fig.~\ref{fig:mb3dp0}(c) corresponds to $h/J=5$ which represents a system in the localized phase.

In Figs.~\ref{fig:mb3dp0}(d)-(f) the same analysis performed for the homogeneous Heisenberg model ($\Delta=1$) is presented for the same initial state.
Even though the localized phase exhibits a behavior akin to the non-interacting case, dynamics of the probabilities in the extended phase and at the transition points are dramatically different.
As shown in Fig.~\ref{fig:mb3dp0}(e) in the transition point where $h_c/J=0.9$ (see Fig.~\ref{fig:mqfih0ND}(b) in the main text), a clear separation between ground- and excited-states is obvious. 
Comparing Figs. \ref{fig:mb3dp0}(b) and (e) at the transition points, we can deduce that by increasing the interaction strength $\Delta$, the particle excitations tend to occupy sites with higher energies, emerging a clear separation between the excited sites and those who remain in the ground state, see Fig.~\ref{fig:mb3dp0}(e).

\section{Appendix C: Many-body QFI scaling}\label{C}
\begin{figure}[h]	
	\includegraphics[width=.49\linewidth]{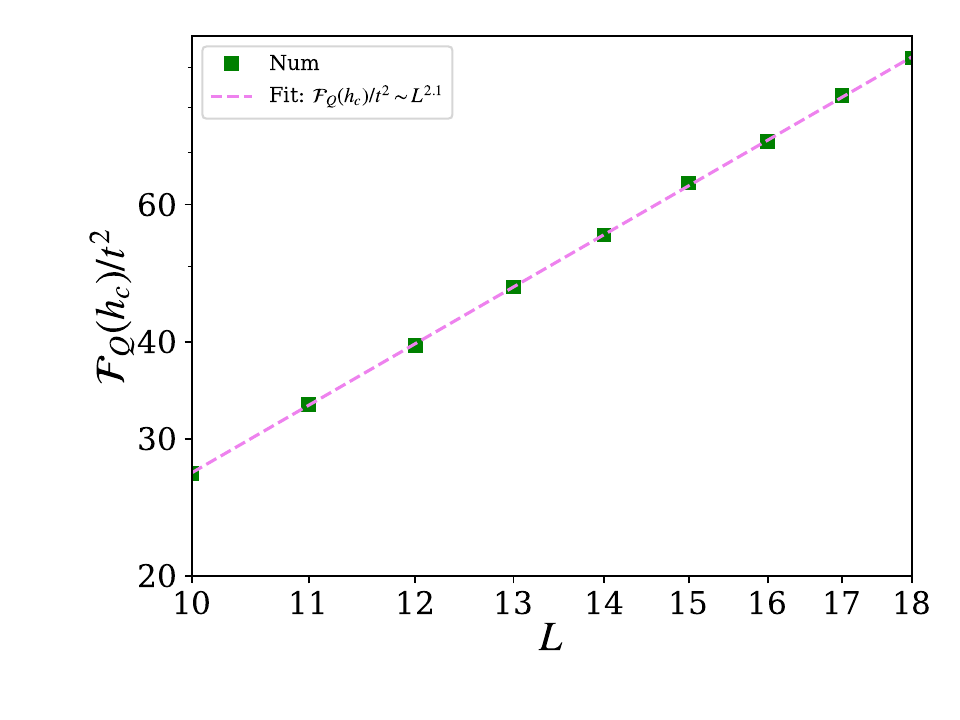}
	\put(-195,178){(a)}
	\includegraphics[width=.49\linewidth]{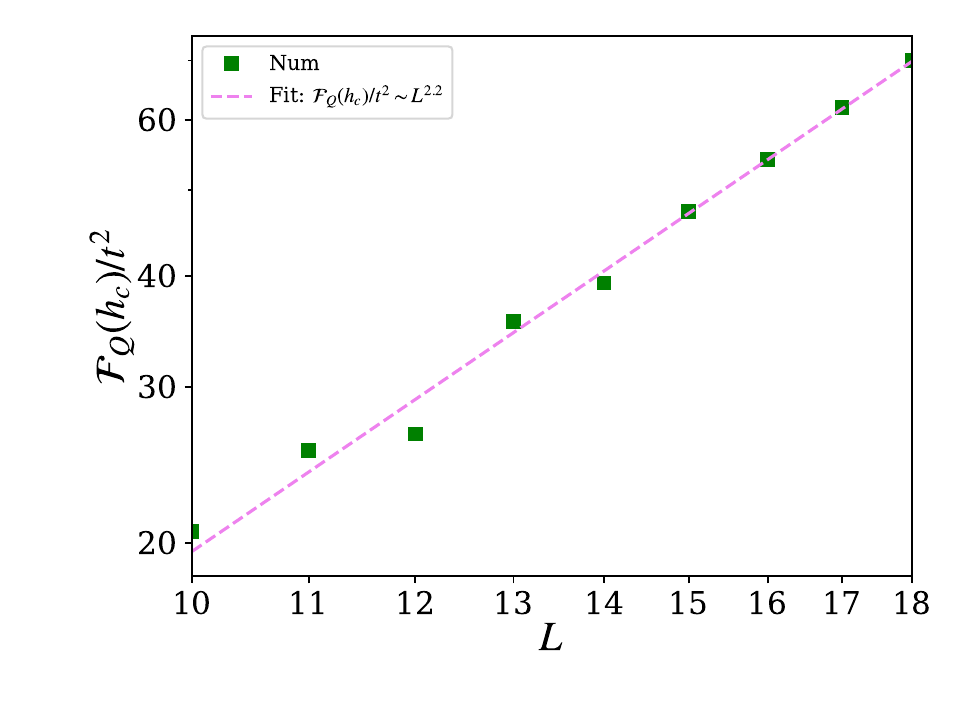}
	\put(-195,178){\small(b)}
	\caption{the normalized QFI at the transition point $\mathcal{F}_Q(h_c)/t^2$ for excitation number $N=3$ as a function of $L$ for (a) $\Delta=0$ and (b) $\Delta=1$. The best fitting exponents for $\mathcal{F}_Q(h_c)/t^2 \sim L^{\beta}$ are $\beta= 2.1$ for $\Delta=0$ and $\beta=2.2$ for $\Delta=1$.}\label{fig:mbQFI}
\end{figure}

In studying the single-particle case, we have shown in the main text that the QFI scales as $\mathcal{F}_Q \sim t^2 L^\beta$ where $\beta$ depends on the phase of the system, which in turn, is determined by $h$ and $L$ [see Fig.~\ref{fig:2}(c) in the main text].
In particular, we find $\beta{=}2$ at the transition point.
Meanwhile, in the many-body case depending on the number of excitations in the initial state we find the scaling function as $\mathcal{F}_Q \sim t^2 L^\beta N^\alpha$.
Unlike $\beta$ which dramatically changes with the phase, $\alpha$ is almost independent of $h$.
This behavior especially is observed near the transition point. Therefore, for a fixed number of excitations the QFI scales as $\mathcal{F}_Q(h_c) \sim t^2 L^\beta$ with $\beta \approx 2$.
To clearly show this point, in Figs. \ref{fig:mbQFI}(a) and (b) for both $\Delta=0$ and $\Delta=1$ we plot the normalized QFI at the transition points $h=h_c$ for a fixed number of excitations ($N{=}3$) and different system sizes $L$.
The fitting lines confirm the power law scaling $\mathcal{F}_Q(h_c)/t^2 \sim L^\beta$ with $\beta= 2.1$ for $\Delta=0$ and $\beta= 2.2$ for $\Delta=1$, which are very close to the value obtained from single-particle case $\beta=2$.

\end{document}